\documentclass[preprint,12pt,number,sort&compress]{elsarticle}
\pdfoutput=1
\usepackage{graphicx}
\usepackage{latexsym,amssymb,amsmath}
\usepackage{bm}
\usepackage[caption=false]{subfig}
\usepackage{xcolor}
\usepackage{mathrsfs}
\begin{document}
\begin{frontmatter}

\title{Structural correlations and dependent scattering mechanism on the radiative properties of random media}
\author{B. X. ~Wang}
\author{C. Y. ~Zhao\corref{cor1}}
\ead{Changying.zhao@sjtu.edu.cn}
\address{Institute of Engineering Thermophysics, Shanghai Jiao Tong University, Shanghai, 200240, People's Republic of China}
\cortext[cor1]{Corresponding author}
\begin{abstract}
The dependent scattering mechanism is known to have a significant impact on the radiative properties of random media containing discrete scatterers. Here we theoretically demonstrate the role of dependent scattering on the radiative properties of disordered media composed of nonabsorbing, dual-dipolar particles. Based on our theoretical formulas for the radiative properties for such media, we investigate the dependent scattering effects, including the effect of modification of the electric and magnetic dipole excitations and the far-field interference effect, both induced and influenced by the structural correlations. We study in detail how the structural correlations play a role in the dependent scattering mechanism by using two types of particle system, i.e., the hard-sphere system and the sticky-hard-sphere system. We show that the inverse stickiness parameter, which controls the interparticle adhesive force and thus the particle correlations, can tune the radiative properties significantly. Particularly, increasing the surface stickiness can result in a higher scattering coefficient and a larger asymmetry factor. The results also imply that in the present system, the far-field interference effect plays a dominant role in the radiative properties while the effect of modification of the electric and magnetic dipole excitations is more subtle. Our study is promising in understanding and manipulating the radiative properties of dual-dipolar random media. 
\end{abstract}
\end{frontmatter}


\section{Introduction}
Studying the radiative properties of micro/nanoscale disordered media is not only of great fundamental importance in understanding the light-matter interaction physics, like Anderson localization \cite{wiersma1997localization,Storzer2006,Segev2013,Sperling2016NJP} and anomalous transport behaviors of radiation (or light) \cite{kopPRL1997,zimnyakovJETPL2005,barthelemyNature2008,bertolottiPRL2010}, but also has profound implications in applications such as random lasers \cite{Cao1999,Wiersma2008}, solar energy harvesting and conversion \cite{Vynck2012,Fang2015JQSRT,Liew2016ACSPh,liuJOSAB2018}, radiative cooling \cite{zhaiScience2017,baoSEMSC2017} and structural color generation \cite{xiaoSciAdv2017}, etc. In such media, radiation is scattered and absorbed in a very complicated way, which is usually described by the radiative transfer equation (RTE) in the mesoscopic scale. The radiative properties entering into RTE, including the scattering coefficient $\kappa_s$, absorption coefficient $\kappa_a$ and phase function $P(\mathbf{\Omega}',\mathbf{\Omega})$ (where $\mathbf{\Omega}'$ and $\mathbf{\Omega}$ denote incident and scattered directions, respectively), depend on the microstructures as well as the permittivity and permeability of the composing materials. Conventionally, for disordered media containing discrete scatterers, the radiative properties are theoretically predicted under the independent scattering approximation (ISA), i.e., in which the discrete inclusions scatter electromagnetic waves independently without any interference effects taken into account \cite{lagendijk1996resonant,VanRossum1998,tsang2004scattering,sheng2006introduction}.

ISA is valid only when the scatterers are far apart from each other and each scatterer scatters light as if no other scatters exist \cite{VanRossum1998,mishchenko2006multiple,tsang2004scattering,sheng2006introduction}. As the concentration of scattering particles in disordered media rises, the scattered waves from different scatterers interfere and ISA fails \cite{garciaPRA2008,wangIJHMT2015,Naraghi2015}. This fact leads to many authors into the considerations on the effect of ``dependent scattering'' in the last a few decades \cite{tien1987thermal,kumar1990dependent,leeJTHT1992,ivezicIJHMT1996,durantJOSAA2007,garciaPRA2008,nguyenOE2013,wangIJHMT2015,Naraghi2015,maJQSRT2017,wangPRA2018}, in order to correctly predict the radiative properties. Generally, the mechanisms of dependent scattering can be classified into two categories. The first category is the recurrent scattering mechanism \cite{Vantiggelen1990JPCM, Cherroret2016}, which denotes the multiple scattering trajectories visiting the same particle more than once and resulting in a closed or half-closed loop. This includes the well-known phenomena such as Anderson localization \cite{wiersma1997localization} and the coherent backscattering cone \cite{mishchenko2006multiple}. The other category is the interference induced by the structural correlations.
Taking the random media composed of hard particles as an example, the finite size of a particle would create a rigid exclusion volume that forbids other particles to penetrate into, which leads to structural correlations in terms of the particle position distribution probability functions \cite{fradenPRL1990,tsang2004scattering2,rojasochoaPRL2004}. The structural correlations will lead to definite phase differences among the scattered waves, which can well preserve over the statistical average procedure. Therefore constructive or destructive interferences among the scattered waves occur and thus affect the transport properties of radiation. This is also called ``partial coherence'' by Lax \cite{laxRMP1951,laxPR1952}. Moreover, when the correlation length of particle positions is comparable with the wavelength, the structural correlations then play a central role in determining the radiative properties \cite{fradenPRL1990,mishchenkoJQSRT1994,rojasochoaPRL2004}.

Generally speaking, the structural correlations are not only affected by the size and packing density (or volume fraction) of the scatterers, but also by the interaction potential between them. Several typical kinds of interaction potential among particles, for example, the pure hard-sphere potential \cite{wertheimPRL1963,fradenPRL1990}, the surface adhesive potential \cite{Baxter1968,Frenkel2002} and the interparticle Coulombic electrostatic potential \cite{rojasochoaPRL2004,bresselJSQRT2013}, can be realized experimentally. By controlling the interaction potential and thus the structural correlations, a flexible manipulation of the radiative properties of random media can be achieved \cite{Peng2007,leseurOptica2016,Froufe-PerezPNAS2017}.  

In this paper, we consider random media consisting of nonabsorbing, dual-dipolar spherical nanoparticles, in which high-order Mie multipolar modes in the particles are negligible and only electric and magnetic dipoles are excited \cite{geffrinNC2012,Gomez-MedinaPRA2012,Zambrana-PuyaltoOL2013,Zambrana-PuyaltoOE2013,schmidtPRL2015}. We aim to comprehensively reveal the dependent scattering effects on the radiative properties, which are induced and influenced by the structural correlations, based on our recently developed rigorous theory \cite{wangPRA2018}. The theory provides analytical expressions for the effective propagation constant, scattering coefficient and phase function for the random media, by means of the multipole expansion method and quasicrystalline approximation (QCA) for the Foldy-Lax equations (FLEs). By investigating two types of particle systems, i.e., the hard-sphere system and the sticky-hard-sphere system, we demonstrate in detail how the structural correlations play a role in the dependent scattering mechanism, including the effect of modification of the electric and magnetic dipole excitations and the far-field interference effect. We show that the inverse stickiness parameter, which controls the interparticle adhesive force and thus the structural correlations, can tune the radiative properties significantly. The results imply that in the present system composed of moderate-refractive-index dual-dipolar particles, the far-field interference effect plays a dominant role in the radiative properties while the effect of modification of the electric and magnetic dipole excitations is more subtle. Our study is promising in understanding and manipulating the radiative properties of dual-dipolar random media. 
\section{Theory}\label{theory}
In this paper, we will consider the radiative properties of a random medium consisting of $N$ identical dual-dipolar particles. In the random media, all the particles are assumed to be isotropic, homogeneous and hard spheres with a radius of $a$. Their positions are regarded as fixed if they are static or move sufficiently slower than the electromagnetic waves \cite{mishchenko2006multiple,mishchenko2014electromagnetic}. Furthermore, the random medium is also supposed to be statistically homogeneous and isotropic. We will not take any quantum or nonlinear effects into account. Under these assumptions, we will briefly summarize the main formulas of this theory to determine radiative properties of such media considering the dependent scattering effects \cite{wangPRA2018}.
\subsection{Effective propagation constant and scattering phase function}\label{theory1}
\begin{figure}[htbp]
	\centering
	\includegraphics[width=0.6\linewidth]{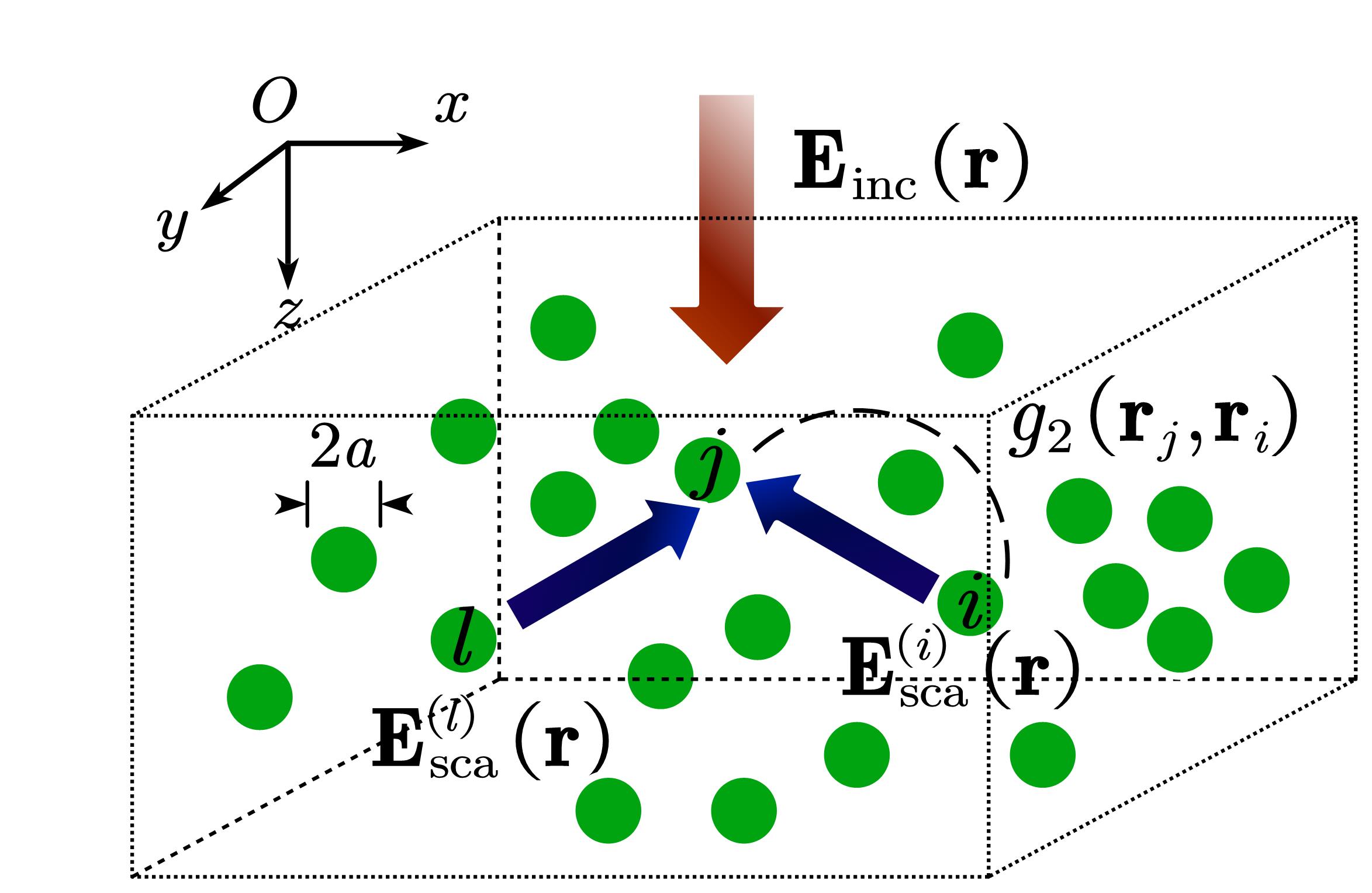}
	\caption{A schematic of Foldy-Lax equations for multiple scattering of electromagnetic waves in randomly distributed spherical particles. The dotted lines stand for an imaginary boundary of the random medium slab. The particles are denoted as $i$, $j$, $l$, etc. The dashed line stands for the pair distribution function (PDF) $g_2(\mathbf{r}_j,\mathbf{r}_i)$ between the $j$-th and $i$-th particle. The thick arrow indicates the propagation direction of the incident wave (which is along the $z$-axis), while the thin arrows stand for the propagation directions of the partial scattered waves from particle $i$ to $j$ and from $l$ to $j$.}
	\label{system_config}
\end{figure}

Following from our assumptions on the random medium, the electromagnetic interaction of the incident light with it is then described by the well-known Foldy-Lax equations (FLEs), which are equivalent to Maxwell equations in terms of multiple scattering of light. The FLEs for $N$ particles are given by \cite{laxRMP1951,varadanJOSAA1985,mackowskiJOSAA1996,tsang2004scattering2}
\begin{equation}\label{fl_eq}
\mathbf{E}_{\text{exc}}^{(j)}(\mathbf{r})=\mathbf{E}_{\text{inc}}(\mathbf{r})+\sum_{i=1\atop i\neq j}^{N}\mathbf{E}_{\text{sca}}^{(i)}(\mathbf{r}),
\end{equation}
where $\mathbf{E}_{\text{inc}}(\mathbf{r})$ is the electric field of the incident radiation, $\mathbf{E}_{\text{exc}}^{(j)}(\mathbf{r})$ is the electric component of the so-called exciting field impinging on the vicinity of the $j$-th particle, and $\mathbf{E}_{\text{sca}}^{(i)}(\mathbf{r})$ is  electric component of partial scattered waves from the $i$-th particle. The schematic of FLEs is shown in Fig.\ref{system_config} .

For spherical particles, it is convenient to expand the electric fields in VSWFs, where the expansion coefficients correspond to multipoles excited by the particles.  The exciting field $\mathbf{E}_{\text{exc}}^{(j)}(\mathbf{r})$ is then expanded as
\begin{equation}\label{exc_expansion}
\mathbf{E}_{\text{exc}}^{(j)}(\mathbf{r})=\sum_{mnp} c_{mnp}^{(j)}\mathbf{N}^{(1)}_{mnp}(\mathbf{r}-\mathbf{r}_j),
\end{equation}
where $\mathbf{N}^{(1)}_{mnp}(\mathbf{r}-\mathbf{r}_j)$ is the regular VSWF, defined in \ref{vswf_appendix}). The summation $\sum_{mnp}$ is abbreviated for $\sum_{n=1}^{\infty}\sum_{m=-n}^{m=n}\sum_{p=1}^{2}$. Here $n$ is set to be 1 because only electric and magnetic dipole modes are taken into account. This is valid when the dual-dipolar particles are not so densely packed that near-field coupling may excite higher order multipoles. The subscript $p=1, 2$ denotes magnetic (TM) or electric (TE) modes, respectively. From the expansion coefficients of the exciting field, the scattering field from the $i$-th particle can be expressed as \cite{mackowskiJOSAA1996,tsang2004scattering}
\begin{equation}\label{sca_expansion}
\mathbf{E}_{\text{sca}}^{(i)}(\mathbf{r})=\sum_{n=1, mp} c_{mnp}^{(i)}T_{np}\mathbf{N}^{(3)}_{m1p}(\mathbf{r}-\mathbf{r}_i),
\end{equation}
where $\mathbf{N}^{(3)}_{mnp}(\mathbf{r}-\mathbf{r}_j)$ is the outgoing VSWF, defined in \ref{vswf_appendix}). For a homogeneous spherical particle the $T$-matrix elements $T_{np}$ are Mie coefficients, i.e., $T_{12}=a_1$ for the electric dipole and $T_{11}=b_1$ for the magnetic dipole, which can be found from standard textbooks \cite{hulst1957,bohrenandhuffman} and are not listed here. Inserting Eqs.(\ref{exc_expansion}) and (\ref{sca_expansion}) into Eq.(\ref{fl_eq}) and using the translation addition theorem for VSWFs at different origins as well as the orthogonal relation of VSWFs with different orders and degrees (see \ref{vswf_appendix}), we can obtain
\begin{equation}\label{fl_eq4}
c_{mp}^{(j)}=a_{mp}^{(j)}+\sum_{i=1\atop i\neq j}^{N}\sum_{\mu q}c_{\mu q}^{(i)}T_{1q}A_{mp\mu q}^{(3)}(\mathbf{r}_j-\mathbf{r}_i),
\end{equation}
where $A_{mp\mu q}^{(3)}(\mathbf{r}_j-\mathbf{r}_i)$ can translate the outgoing VSWFs centered at $\mathbf{r}_i$ of degree $\mu$ and polarization $q$ to regular VSWFs centered at $\mathbf{r}_j$ of degree $m$ and polarization $p$. Here $n=1$ in the subscript of $c_{mnp}^{(j)}$ and $a_{mnp}^{(j)}$ is omitted. $a_{mp}^{(j)}$ is the expansion coefficient of the incident wave in terms of regular VSWFs centered at $\mathbf{r}_j$. Since the particles can be seen as fixed, the electromagnetic response after a long period of time or over a sufficient large spatial range is computed by taking average of all possible configurations of particle distributions \cite{mishchenko2006multiple,mishchenko2014electromagnetic}. The ensemble average of Eq.(\ref{fl_eq4}) with respect to a fixed particle centered at $\mathbf{r}_j$ is given by
\begin{equation}\label{fl_avg_eq}
\langle c_{mp}^{(j)}\rangle_j=\langle a_{mp}^{(j)}\rangle_j+\Big\langle\sum_{i=1\atop i\neq j}^{N}\sum_{\mu q}c_{\mu q}^{(i)}T_{1q}A_{mp\mu q}^{(3)}(\mathbf{r}_j-\mathbf{r}_i)\Big\rangle_j,
\end{equation}
where $\langle \cdot \rangle_j$ represents the ensemble average procedure with $\mathbf{r}_j$ fixed. For statistically homogeneous random media, ensemble average procedure restores the translational symmetry. The ensemble-averaged electromagnetic field in random media, namely, the coherent field, is a plane wave as proved by Lax \cite{laxPR1952}. Here we only consider transverse electromagnetic wave propagation and assume the random medium only supports a transverse coherent mode with an effective propagation wave vector $\mathbf{K}$. Therefore, the effective exciting field for particle $j$, which is equal to the total coherent field minus the field scattered by the investigated scatterer $j$, takes the following form \cite{laxPR1952}:
\begin{equation}\label{aprox1}
\langle c_{mp}^{(j)}\rangle_j\approx C_{mp}\exp({i\mathbf{K}\cdot\mathbf{r}_j}),
\end{equation}
where $C_{mp}$ is the expansion coefficient of effective exciting wave amplitude at the origin \cite{laxRMP1951,laxPR1952}. $C_{mp}$ only depends on the overall property of the random media. Note under ISA, $C_{mp}$ is always equal to unity. In this spirit, $C_{mp}$ thus quantifies the modification of electric and magnetic dipole excitations. Furthermore the QCA, which expresses high-order correlations among three or more particles using two-particle statistics, is introduced as
\begin{equation}\label{aprox2}
\langle c_{mp}^{(i)}\rangle_{ij}\approx\langle c_{mp}^{(i)}\rangle_{i}\approx C_{mp}\exp({i\mathbf{K}\cdot\mathbf{r}_i}),
\end{equation}
where $\langle \cdot \rangle_{ij}$ denotes the ensemble average procedure with $\mathbf{r}_j$ and $\mathbf{r}_i$ fixed simultaneously. This approximation also amounts to neglecting the fluctuation of the effective exciting field impinging on the \textit{i}-th particle due to a shift of the \textit{j}-th particle from its average position. Strictly speaking, it is valid for a periodic or crystalline medium, and to some extent it is a reasonable approximation for a densely packed medium possessing some partial order (short-range order). QCA was initially developed for both quantum and classical waves \cite{laxPR1952}, and further verified by well-controlled experiments as well as numerical simulations \cite{westJOSAA1994,Nashashibi1999,Siqueira2000}. It is widely used in the prediction of radiative transport properties of disordered materials for applications in remote sensing \cite{liangIEEETGRS2008} as well as thermal radiation transfer \cite{prasherJAP2007,wangIJHMT2015}. 

Without loss of generality, here we assume an incident plane wave linearly polarized over the $y$ axis and propagating along the $z$-axis as shown in Fig.\ref{system_config}, with an amplitude of unity. Inserting Eqs.(\ref{aprox1}-\ref{aprox2}) into Eq.(\ref{fl_avg_eq}), after some manipulations, we are finally able to obtain the following equations containing three unknowns, $K$, $C_{11}$ and $C_{12}$ as
\begin{equation}\label{c_eq1}
\begin{split}
(1-n_0\tilde{A}_{1111}(K)b_1)C_{11}-n_0\tilde{A}_{1211}(K)a_1C_{12}=0,
\end{split}
\end{equation}
\begin{equation}\label{c_eq2}
\begin{split}
(1-n_0\tilde{A}_{1111}(K)a_1)C_{12}-n_0\tilde{A}_{1211}(K)b_1C_{11}=0.
\end{split}
\end{equation}
In this circumstance, $C_{-11}=-C_{11}$, $C_{01}=C_{02}$ and $C_{-12}=-C_{12}$. Eliminating $C_{11}$ and $C_{12}$, therefore we get the final dispersion relation for the effective propagation constant as
\begin{equation}\label{dispersion_relate_eq}
[1-n_0\tilde{A}_{1111}(K)b_1][1-n_0\tilde{A}_{1111}(K)a_1]-n_0^2\tilde{A}_{1211}^2a_1b_1=0,
\end{equation}
where the effective propagation constant $K$ can be solved in the upper complex plane. The elements of $\tilde{\mathbf{A}}(\mathbf{K})$ appearing in the above equations are obtained using the following relations:
\begin{equation}
\begin{split}
&\tilde{A}_{m1\mu 1}(\mathbf{K})=\tilde{A}_{m2\mu 2}(\mathbf{K})=\sqrt{\frac{(1-\mu)!(1+m)!}{(1+\mu)!(1-m)!}}(-1)^{m}\\&\cdot\sum_{n}a(\mu,1|-m,1|n)a(1,1,n)I_{n}^{m,\mu}(\mathbf{K}),
\end{split}
\end{equation}
\begin{equation}
\begin{split}
&\tilde{A}_{m2\mu 1}(\mathbf{K})=\tilde{A}_{m1\mu 2}(\mathbf{K})=\sqrt{\frac{(1-\mu)!(1+m)!}{(1+\mu)!(1-m)!}}(-1)^{m+1}\\&\cdot\sum_{n}a(\mu,1|-m,1|n,n-1)b(1,1,n)I_{n}^{m,\mu}(\mathbf{K}),
\end{split}
\end{equation}
where 
\begin{equation}
\begin{split}
I_{n}^{m,\mu}(\mathbf{K})=\int_0^\infty 4\pi(-i)^nh_n(kr)j_{n}(Kr)Y_{n}^{\mu-m}(\theta_K,\phi_K)g_2(r)r^2dr
\end{split}
\end{equation}
is an integral involving the PDF, and can be evaluated numerically for a known $g_2(r)$. Note throughout this paper, all the integrals, if not specified, are performed over the entire real (for position vector) or reciprocal (for reciprocal vector) spaces. Here $Y_n^{m}(\theta,\phi)$ are spherical harmonics, and coefficients that contain $m, \mu,n$ including $a(m,1|-\mu,1|n)$, $a(\mu,1|-m,1|n)$, $a(1,1,n)$, $b(1,1,n)$ are related to Wigner 3-$j$ symbols and listed in \ref{vswf_appendix}. 

An additional relation is necessary to solve $C_{12}$ and $C_{11}$ on the basis of Eqs.(\ref{c_eq1}) and (\ref{c_eq2}). This can be done by considering the relationship between the effective propagation constant and the transmission coefficient of the coherent field. According to Ref.\cite{wangPRA2018}, we obtain
\begin{equation}\label{extionction_theorem}
K^2-k^2=\frac{6\pi in_0}{k}(a_1C_{12}+b_1C_{11}).
\end{equation}
By combining Eqs.(\ref{c_eq1}),(\ref{dispersion_relate_eq}) and (\ref{extionction_theorem}), the effective exciting field amplitudes $C_{12}$ and $C_{11}$ can be solved, which allow us to determine how the dependent scattering effects, induced by the structural correlations, play a role in the modification of the electric and magnetic dipole excitations.

After calculating the effective propagation constant for the coherent wave, we  compute the scattering intensity to derive the scattering phase function defined for incoherent waves \cite{maAO1988}. The ensemble averaged total intensity is given by
\begin{equation}
I(\mathbf{r})=\langle\mathbf{E}(\mathbf{r})\mathbf{E}^*(\mathbf{r})\rangle=\langle[\mathbf{E}_{\text{inc}}(\mathbf{r})+\mathbf{E}_\text{s}(\mathbf{r})]\cdot[\mathbf{E}_{\text{inc}}^*(\mathbf{r})+\mathbf{E}_\text{s}^*(\mathbf{r})]\rangle.
\end{equation}
where $\mathbf{E}(\mathbf{r})=\mathbf{E}_{\text{inc}}(\mathbf{r})+\mathbf{E}_\text{s}(\mathbf{r})$ is the total field, and $\mathbf{E}_\text{s}$ is the total scattered field generated by all the particles. The superscript $*$ denotes the complex conjugate. The coherent intensity is defined as
\begin{equation}
I_{\text{coh}}(\mathbf{r})=\mathbf{E}_{\text{coh}}(\mathbf{r})\mathbf{E}_{\text{coh}}^*(\mathbf{r})
\end{equation}
where $\mathbf{E}_{\text{coh}}(\mathbf{r})$ is the coherent field that is the ensemble averaged total field $ \mathbf{E}_{\text{coh}}(\mathbf{r})=\langle\mathbf{E}(\mathbf{r})\rangle=\langle\mathbf{E}_{\text{inc}}(\mathbf{r})+\mathbf{E}_\text{s}(\mathbf{r})\rangle$. Therefore the incoherent intensity, defined as the difference between total intensity and coherent intensity, is given by
\begin{equation}
I_{\text{ich}}(\mathbf{r})=I(\mathbf{r})-I_\text{coh}(\mathbf{r})=\langle\mathbf{E}_\text{s}(\mathbf{r})\mathbf{E}_\text{s}^*(\mathbf{r})\rangle-\langle\mathbf{E}_\text{s}(\mathbf{r})\rangle\langle\mathbf{E}_\text{s}^*(\mathbf{r})\rangle,
\end{equation}
The incoherent intensity denotes the light intensity generated by random fluctuations of the medium, which is also called diffuse intensity by some authors \cite{mackowskiJQSRT2013}. By expressing the total scattered wave in VSWFs, applying the far-field and on-shell approximations as well as Fourier transform technique, we can finally derive the following expression for the incoherent intensity \cite{wangPRA2018}:
\begin{equation}\label{incoh_sca_eq7}
\begin{split}
I_{\text{ich}}(\mathbf{r})&=n_0\sum_{mpm'p'}\iiint d\hat{\mathbf{p}}d\mathbf{p}'d\mathbf{s}[1+n_0(2\pi)^3H(\mathbf{p}'-\mathbf{p})]T_{1p}T_{1p'}^*C_{mp}C_{m'p'}^* \\&\times  \mathbf{N}^{(3)}_{m1p}(\mathbf{r}+\mathbf{s}/2)\mathbf{N}^{(3)*}_{m1p}(\mathbf{r}-\mathbf{s}/2)\exp(-i\mathbf{p}'\times \mathbf{s})\langle\mathbf{E}(\hat{\mathbf{p}}K)\mathbf{E}^*(\hat{\mathbf{p}}K)\rangle,
\end{split}
\end{equation}
where  $H(\mathbf{p}'-\mathbf{p})$ is the Fourier transform of pair correlation function (PCF) $h_2(\mathbf{r})=g_2(\mathbf{r})-1$ as
\begin{equation}\label{Hq_eq}
H(\mathbf{p}'-\mathbf{p})=\frac{1}{(2\pi)^3}\int d\mathbf{r}h_2(r)\exp{[-i(\mathbf{p}'-\mathbf{p})\cdot\mathbf{r}]}.
\end{equation}
The physical significance of Eq.(\ref{incoh_sca_eq7}) is that the incoherent intensity arises from the process in which the total intensity propagating along $\hat{\mathbf{p}}$ is scattered into the direction $\hat{\mathbf{p}}'$, and the total incoherent intensity should be integrated over all possible incident and scattering directions. It is the process that is described by the conventional RTE \cite{mishchenko2006multiple}. Therefore, the quantity in the integral is indeed the different scattering coefficient, which is given by
\begin{equation}\label{pf_qca}
\begin{split}
\frac{d\kappa_\mathrm{s}}{d\varOmega_\text{s}}&=n_0\sum_{mpm'p'}\int d\mathbf{s}[1+n_0(2\pi)^3H(\mathbf{p}'-\mathbf{p})]T_{1p}T_{1p'}^*C_{mp}C_{m'p'}^*\\&\times  \mathbf{N}^{(3)}_{m1p}(\mathbf{r}+\mathbf{s}/2)\mathbf{N}^{(3)*}_{m1p}(\mathbf{r}-\mathbf{s}/2)\exp(-i\mathbf{p}'\cdot \mathbf{s}),
\end{split}
\end{equation}
where $\varOmega_\text{s}$ indicates the scattering solid angle defined as the angle between incident direction $\mathbf{p}$ and scattering direction $\mathbf{p}'$. By utilizing the asymptotic property for VSWFs in the far field (see \ref{far-field_appendix}), above equation can be calculated as \cite{wangPRA2018}
\begin{equation}\label{pf_qca2}
\begin{split}
\frac{d\kappa_\mathrm{s}}{d\theta_\text{s}}&=\frac{9n_0}{4k^2}[1+n_0(2\pi)^3H(\mathbf{p}'-\mathbf{p})]\Big[|a_1C_{21}\pi_n(\cos\theta_\text{s})+b_1C_{11}\tau_n(\cos\theta_\text{s})|^2\\
&+|b_1C_{11}\pi_n(\cos\theta_\text{s})+a_1C_{21}\tau_n(\cos\theta_\text{s})|^2\Big],
\end{split}
\end{equation}
where $\theta_\text{s}$ is the polar scattering angle, and the dependency on azimuth angle is integrated out. The functions $\tau_n(\cos\theta_\text{s})$ and $\pi_n(\cos\theta_\text{s})$ are defined in \ref{far-field_appendix}. Under the on-shell and far-field approximations \cite{wangPRA2018}, the argument in the Fourier transform of the PCF is given by $|\mathbf{p}'-\mathbf{p}|=\sqrt{K^2+k^2-2Kk\cos\theta_s}$.

Eq.(\ref{pf_qca2}) is the main formula providing the scattering coefficient and phase function with considerations on the dependent scattering effects. Accordingly, the structural correlations among particles, i.e., $h_2(r)$, induce and influence the dependent scattering mechanism in two ways. The first is manifested in the structure factor defined as $S(\mathbf{q})=1+n_0(2\pi)^3H(\mathbf{p}'-\mathbf{p})$ where $\mathbf{q}=\mathbf{p}'-\mathbf{p}$. The structure factor is widely used by many authors as the first order dependent-scattering correction to the differential scattering coefficient of ISA, for instance, Refs.\cite{fradenPRL1990,mishchenkoJQSRT1994,rojasochoaPRL2004,yamadaJHT1986,conleyPRL2014,liuJOSAB2018}, which describes the far-field interference between first-order scattered waves of different particles, also named as the interference approximation (ITA) by some authors \cite{dickJOSAA1999}. The second role of the structural correlations is to introduce the effective exciting field amplitudes $C_{12}$ and $C_{11}$, which are not accounted by either ISA or ITA. If they are substantially different from unity, the scattering coefficient and phase function will also be affected. In Section \ref{results}, we will discuss in detail the two effects induced and influenced by the structural correlations.

Note our dependent scattering model does not account for the recurrent scattering effect \cite{Aubry2014PRL}, which indicate the multiple scattering trajectories visiting the same particle more than once and resulting in a closed loop. This effect is significant for extremely strong scattering media, for instance, cold atomic clouds \cite{Cherroret2016}, and can occur even when there are no structural correlations \cite{vanTiggelenPRE2006}. However, since we only focus on moderately scattering media and the dependent scattering effects due to the structural correlations, the recurrent scattering effect is beyond the scope of this paper.

\subsection{Pair correlation function}
By now the undetermined quantities in the theoretical formulas are the PCF $h_2(r)$ and its Fourier transform $H(\mathbf{q})$. The PCF directly stands for two-particle statistics in the structural correlations. In order to explore how a manipulation of structural correlations can induce and influence the dependent scattering mechanism and thus the radiative properties, in this section, we consider two kinds of disordered particle systems with different structural correlations, including the hard-sphere (HS) system \cite{wertheimPRL1963} and the sticky-hard-sphere (SHS) system \cite{Baxter1968}. 

For the HS system, all particles are randomly distributed and the only restriction for their positions is that they do not be deform by or penetrate each other. The interacting potential is thus given by 
\begin{equation}
U_{\mathrm{HS}}(\mathbf{r})=\begin{cases}
\infty &{0<r<d}\\
0 &{r\geq d}
\end{cases}
\end{equation}
where $d=2a$ is the diameter of the sphere. Inserting this potential into the well known Ornstein-Zernike integral equation and using the Percus-Yevick approximation \cite{wertheimPRL1963}, a closed form solution of the PCF can be obtained. The Fourier transform of the PCF is solved and given by Refs.\cite{wertheimPRL1963,tsang2004scattering2} as
\begin{equation}
H(\mathbf{q})=\frac{F(q)}{1-n_0(2\pi)^3F(q)},
\end{equation}
where
\begin{equation}
\begin{split}
n_0(2\pi)^3F(q)=&24f_v[\frac{\alpha+\beta+\delta}{u^2}\cos u-\frac{\alpha+2\beta+4\delta}{u^3}\sin u\\&-2\frac{\beta+6\delta}{u^4}\cos u+\frac{2\beta}{u^4}+\frac{24\delta}{u^5}\sin u+\frac{24\delta}{u^6}(\cos u-1)]
\end{split}
\end{equation}
with $q=|\mathbf{q}|$, $u=2qa$, $\alpha=(1+2f_v)^2/(1-f_v)^4$, $\beta=-6f_v(1+f_v/2)^2/(1-f_v)^4$, $\delta=f_v(1+2f_v)^2/[2(1-f_v)^2]$.
By taking the inverse Fourier transform, we are able to obtain the PCF for this system as
\begin{equation}\label{ftransform}
h_2(\mathbf{r})=h_2(r)=\int_{-\infty}^{\infty}H(\mathbf{p})\exp{(i\mathbf{p}\cdot\mathbf{r})}d\mathbf{r}
\end{equation}
This model is capable to reproduce the position relations between pairs of spherical particle analytically with a high accurateness \cite{tsang2004scattering2}.

On the other hand, the SHS system is described by the following attractive potential as \cite{tsang2004scattering,Frenkel2002,Baxter1968}, 
\begin{equation}
U_{\mathrm{SHS}}(\mathbf{r})=\begin{cases}
\infty &{0<r<s}\\
\ln{[\frac{12\tau (d-s)}{d}]} & {s<r<d}\\
0 &{r>d}
\end{cases}
\end{equation}
where $\tau$ is a parameter whose inverse $\tau^{-1}$ measures the strength of particle surface adhesion (which will be called the inverse stickiness parameter later on). $(d-s)$ stands for the range of potential, which is assumed to be infinitesimal because this potential is confined on particle surface \cite{Baxter1968}. Again, the Percus-Yevick approximation is utilized in the O-Z equation to solve the PCF of the SHS system using the factorization method of Baxter \cite{Baxter1968,tsang2004scattering2}.  Similarly, the PCF in reciprocal space is solved from
\begin{equation}
\begin{split}
[1+n_0(2\pi)^3H(\mathbf{q})]^{-1}=&\left\{\frac{f_v}{1-f_v}\left[\left(1-tf_v+\frac{3f_v}{1-f_v}\right)\Phi(y)+[3-t(1-f_v)]\Psi(y)\right]+\cos y\right\}^2\\
&+\left\{\frac{f_v}{1-f_v}[y\Phi(y)]+\sin y\right\}^2,
\end{split}
\end{equation}
where $y=qa$, $\Psi(y)=3(\sin y/y^3-\cos y/y^2)$ and $\Phi(y)=\sin y/y$. And the parameter $t$ satisfies the following equation for a given $f_v$ and $\tau$ \cite{tsang2004scattering2}
\begin{equation}
\frac{f_v}{12}t^2-(\tau+\frac{f_v}{1-f_v})t+\frac{1+f_v/2}{(1-f_v)^2}=0
\end{equation}
Therefore, by taking the inverse Fourier transform for $H(\mathbf{q})$, the same as Eq. (\ref{ftransform}), the PCF for sticky-sphere system is also obtained.

\begin{figure}[htbp]
	\centering
	\includegraphics[width=0.6\linewidth]{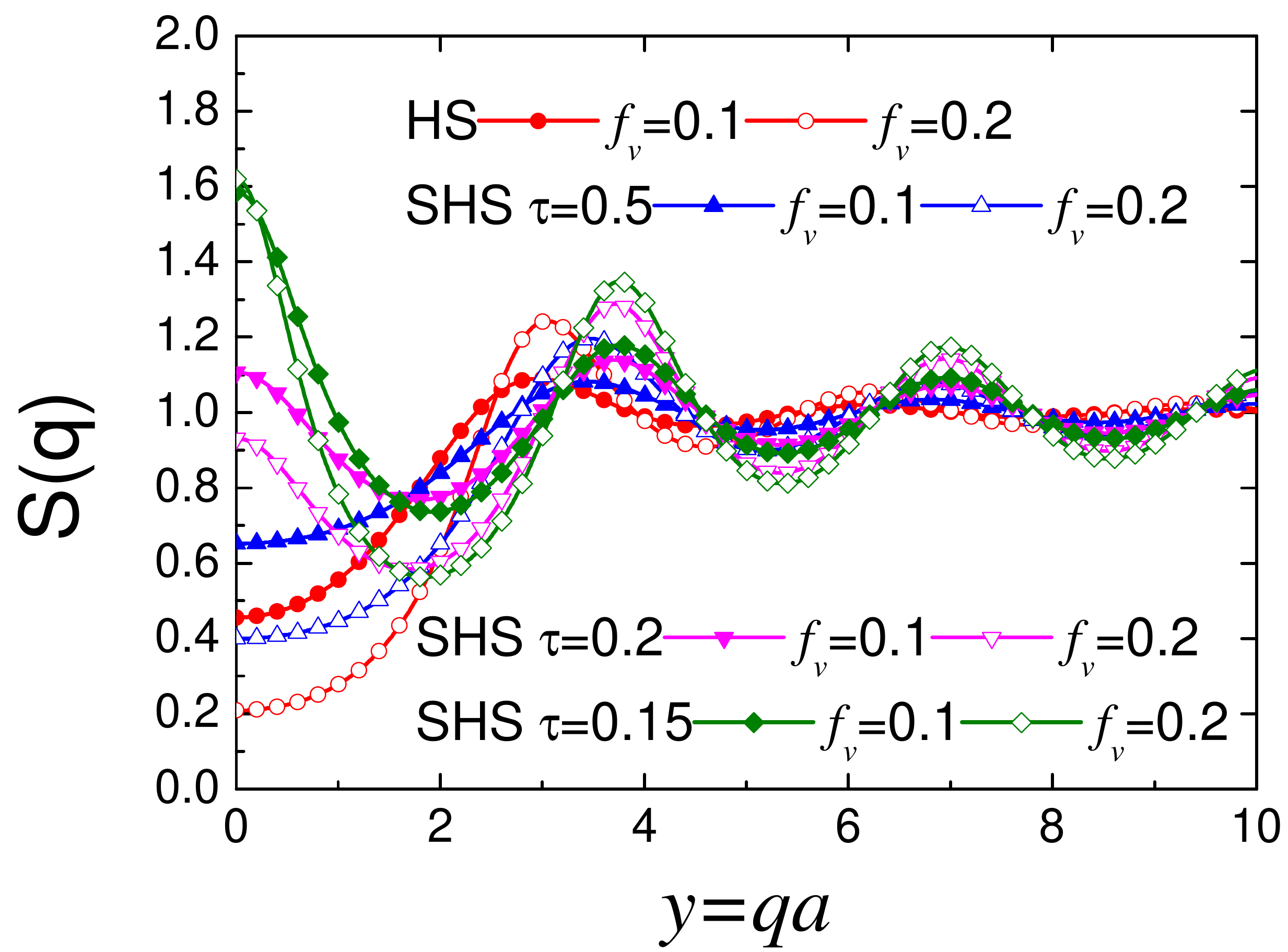}
	\caption{The structure factor $S(q)=1+n_0(2\pi)^3H(\mathbf{q})$ as a function of $y=qa$ for different random systems with different inverse stickiness parameters for particle volume fractions $f_v=0.1$ and $f_v=0.2$.}
	\label{sq}
\end{figure}

Since the structure factor accounts for the far-field interference effect, in Fig.\ref{sq}, we show the structure factor $S(q)=1+n_0(2\pi)^3H(\mathbf{q})$ of different random systems as a function of $y=qa$ for particle volume fractions $f_v=0.1$ and $f_v=0.2$, to aid the analysis in the following section.
\section{Results and Discussions}\label{results}
\begin{figure}[htbp]
	\centering
	\includegraphics[width=0.6\linewidth]{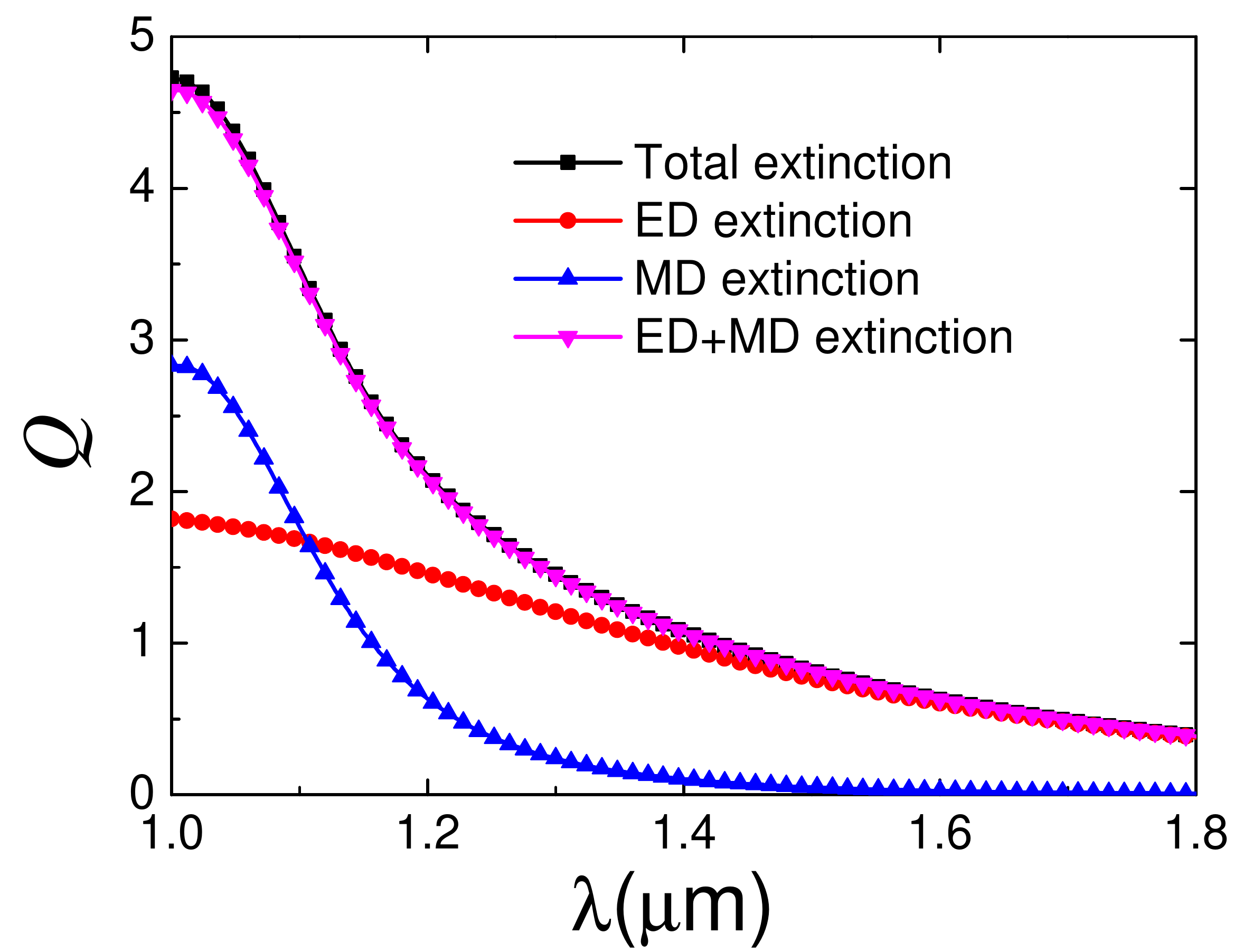}
	\caption{Extinction efficiency $Q=C_{\mathrm{ext}}/(\pi a^2)$ for a single spherical nanoparticle with radius a = 230 nm as a function of the wavelength $\lambda$, where the contributions of electric dipole (ED) and magnetic dipole (MD) as well as the sum of them (ED+MD) are also shown. Here $C_{\mathrm{ext}}$ is the extinction cross section of a single spherical particle calculated from the Mie theory.}
	\label{qext}
\end{figure}

Based on the theoretical formulas on the radiative properties as well as the analytical expressions of HS and SHS PCFs, in this section, we present the obtained results and investigate the dependent scattering effects induced and influenced by the structural correlations. Here we consider a random medium consisting of nonabsorbing zirconia nanoparticles, whose refractive index is set to be $n=2.1$. Porous coatings made of zirconia nanoparticles are widely used to provide high temperature thermal insulation, for which the protection over thermal conduction and radiation heat transfer is of paramount importance \cite{gongSCT2006,martheSCT2013,wangIJHMT2015,wangIJTS2017,chenJQSRT2018}. Here we fix the radius $a$ of the nanoparticle to be $0.23\mathrm{\mu m}$ and investigate the spectral response in the range of $1\le\lambda\le 1.8\mathrm{\mu m}$, in which the electric and magnetic dipoles in the particles are excited while high-order Mie multipolar modes are negligible. This is demonstrated in the extinction efficiency spectra shown in Fig.\ref{qext}, where total the contributions of electric dipole (ED) and magnetic dipole (MD) as well as the sum of them (ED+MD) are also plotted for comparison. 
\subsection{Asymmetry factor and phase function}
\begin{figure}[htbp]
	\centering
	\subfloat{	
		\label{gtau01}
		\includegraphics[width=0.46\linewidth]{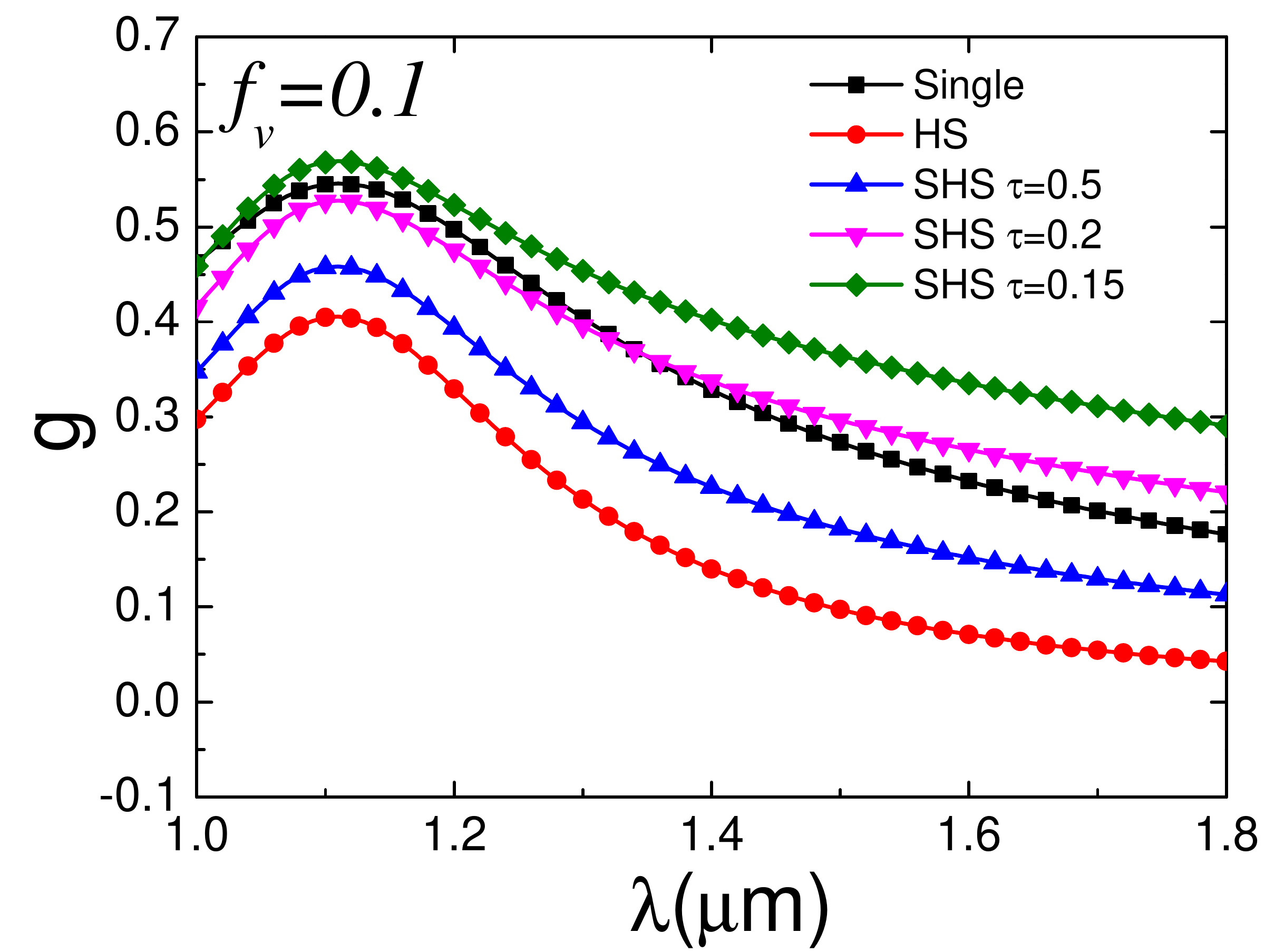}
		
	}
	\hspace{0.01in}
	\subfloat{	
		\label{gtau02}
		\includegraphics[width=0.46\linewidth]{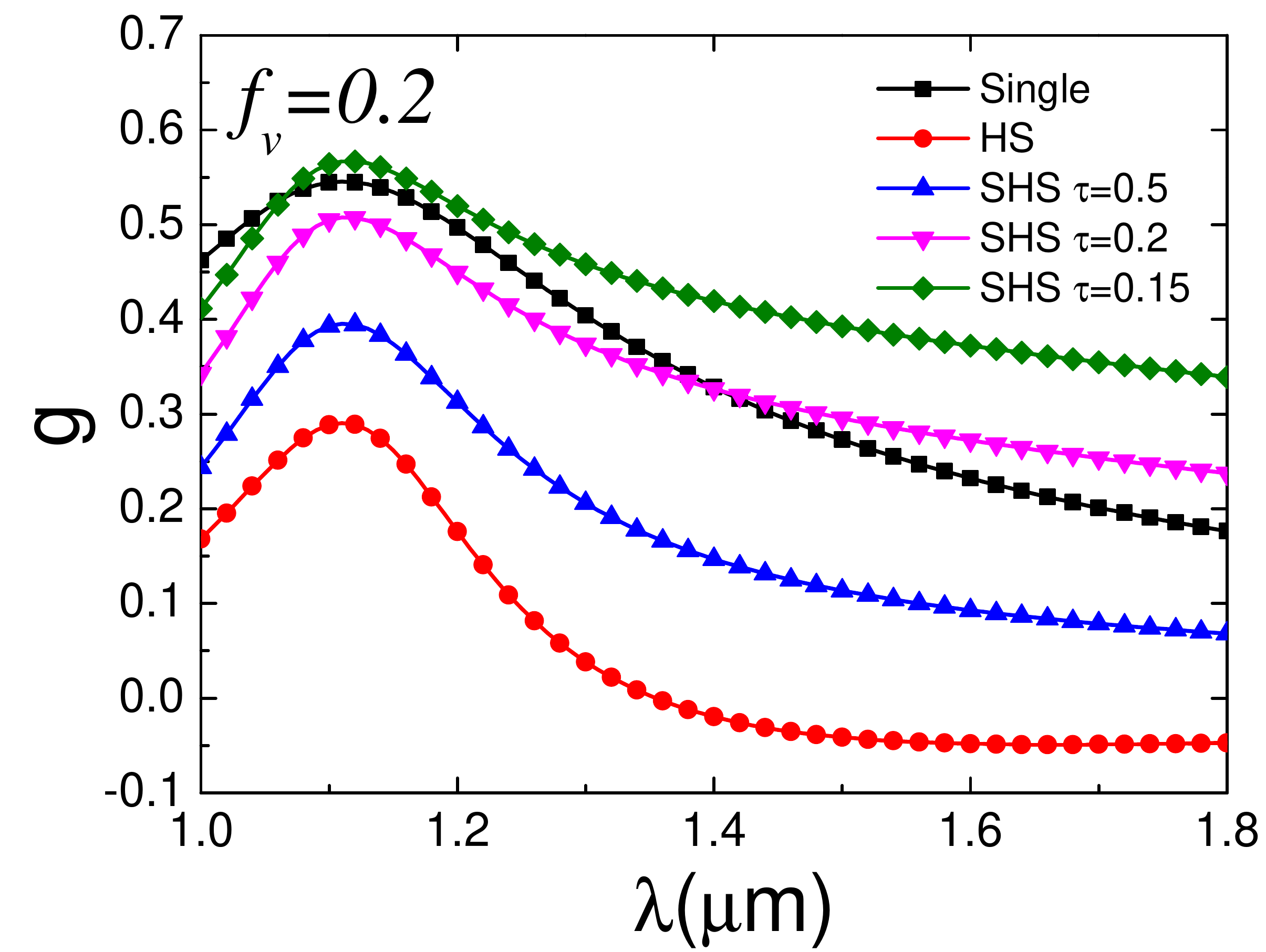}
	}
	\caption{The asymmetry factor $g$ for random systems as a function of wavelength for random media with different stickiness parameters $\tau$. (a) $f_v=0.1$; (b) $f_v=0.2$.}\label{gtau}
\end{figure}
We first study the dependent scattering effects on the asymmetry factor as well as the phase function of the random media. The asymmetry factor as a function of the thermal radiation wavelength for the particle volume fraction $f_v=0.1$ and $f_v=0.2$ is shown in Figs.\ref{gtau01} and \ref{gtau02} respectively. The comparison between different volume fractions will provide information on how the packing density influences the radiative properties. 

It can be observed that the particle correlation affects the asymmetry factor substantially. For the HS systems, the asymmetry factor is much smaller than the single particle case, and for $f_v=0.2$, in the wavelength range of $1.3\le\lambda\le 1.8\mathrm{\mu m}$, we even obtain a negative asymmetry factor. This result is due to the suppression of forward scattering and the enhancement of backscattering in the HS systems. It is well known that randomly packed hard particles with no additional interparticle forces will show a forward scattering suppression \cite{mishchenkoJQSRT1994} where the structure factor is much smaller than unity in the forward scattering direction (i.e., $q\sim0$) according to Fig.\ref{sq}. This suppression becomes more significant when the volume fraction (packing density) is increased as shown in Fig.\ref{sq}. Moreover, if the size parameter $a/\lambda$ is in an appropriate range, a backscattering enhancement can be achieved, as seen in Fig.\ref{sq}, where the structure factor in the backscattering direction is larger than unity (the normalized parameter in the exact backscattering direction $y_{\mathrm{back}}\approx2ka\gtrsim 2$), and this is the case for the present HS systems. This fact is further demonstrated in the phase function $P(\theta_\mathrm{s})$ as a function of the scattering angle $\theta_\mathrm{s}$ at $\lambda=1\mathrm{\mu m}$, as shown in Fig.\ref{gtau}. 

\begin{figure}[htbp]
	\centering
	\subfloat{	
		\label{fv01wavl1phasefunc}
		\includegraphics[width=0.46\linewidth]{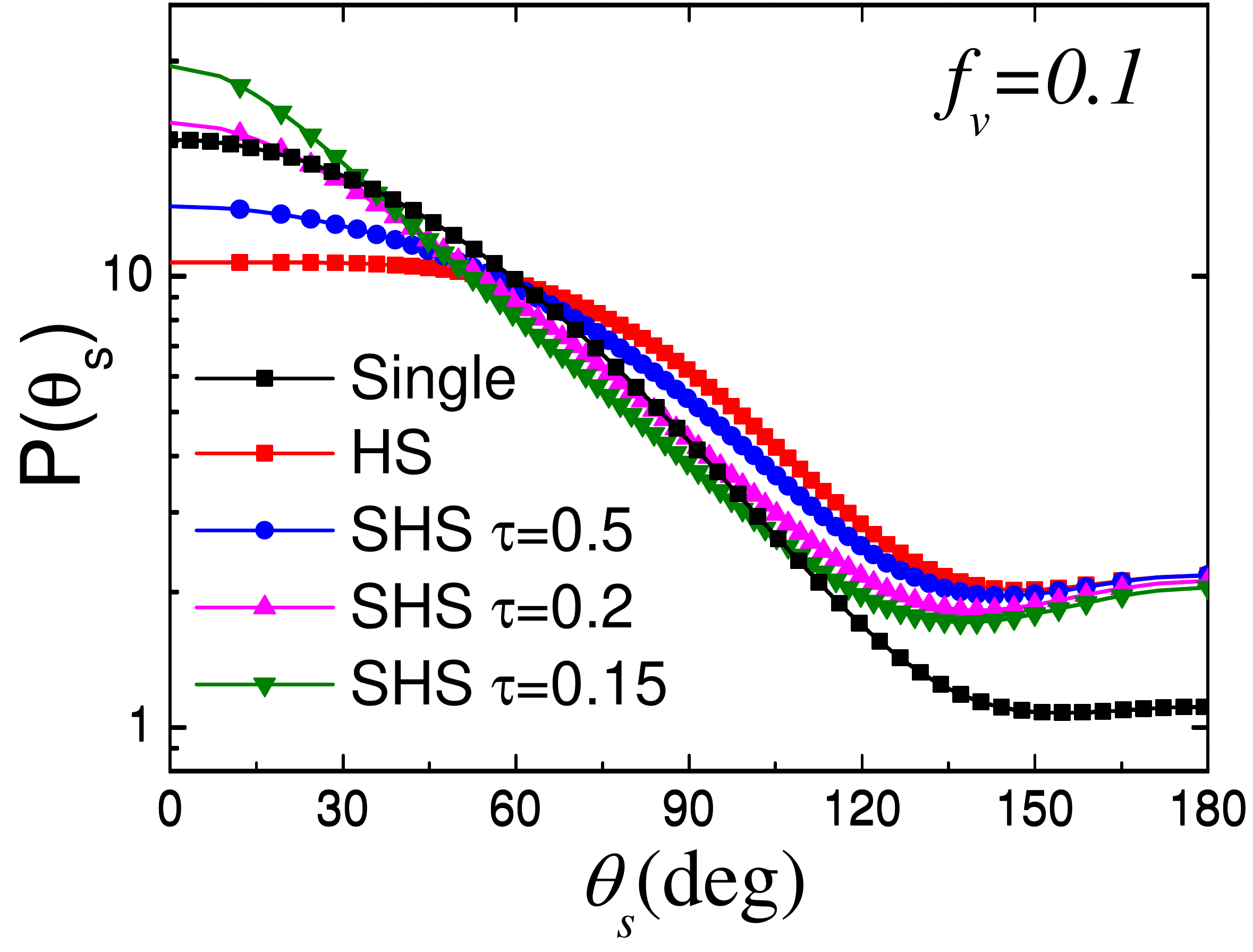}
		
	}
	\hspace{0.01in}
	\subfloat{	
		\label{fv02wavl1phasefunc}
		\includegraphics[width=0.46\linewidth]{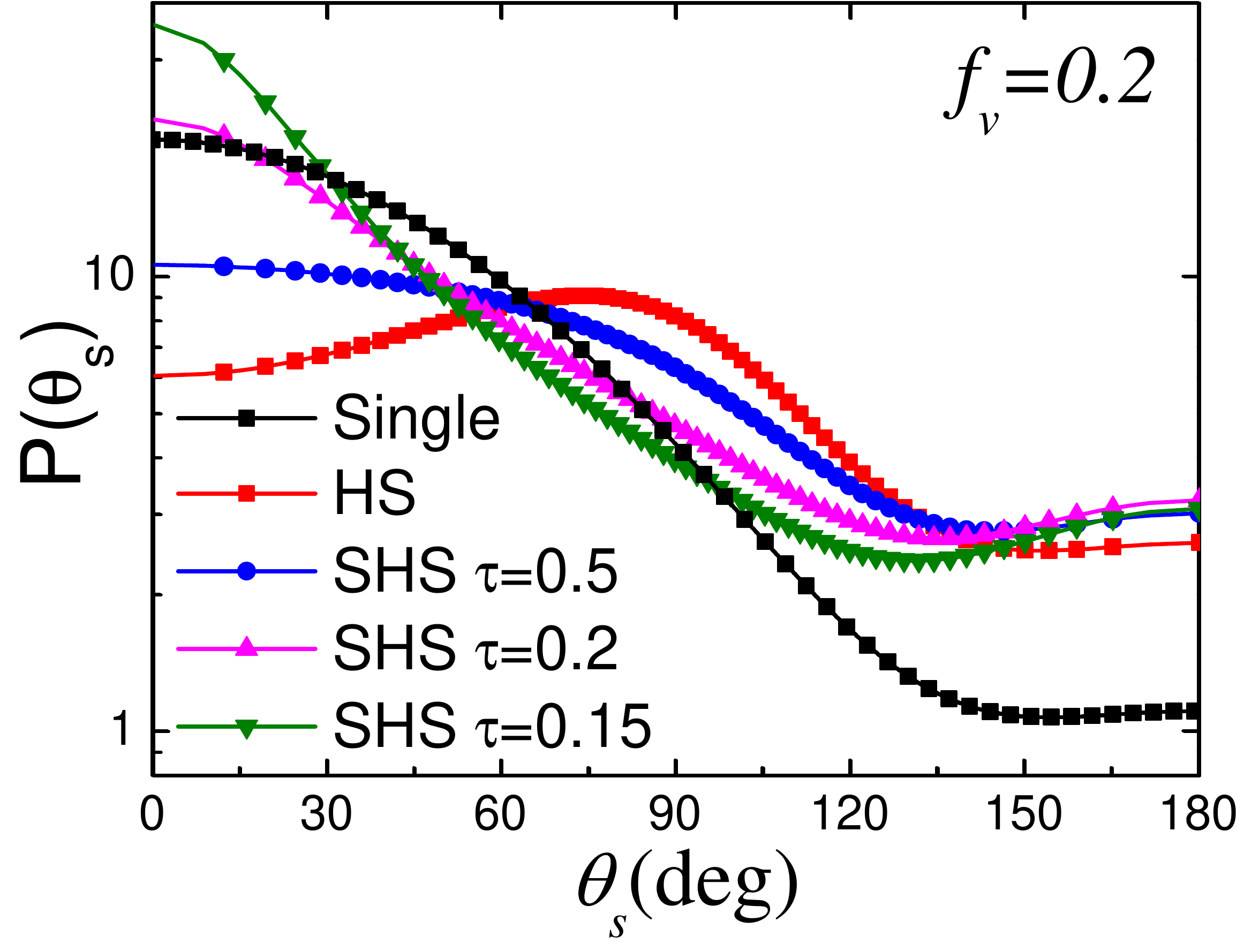}
	}
	\caption{The phase function $P(\theta_\mathrm{s})$ as a function of the scattering angle $\theta_\mathrm{s}$ at $\lambda=1\mathrm{\mu m}$ for random systems with different inverse stickiness parameters. (a) $f_v=0.1$; (b) $f_v=0.2$}\label{phasefunc}
\end{figure}

However, when the inverse stickiness parameter $\tau$ reduces (which indicates a stronger interparticle adhesive force), the asymmetry factor monotonously grows, irrespective of the wavelength. For the SHS systems with $\tau=0.15$, the asymmetry factor even exceeds that of the single particle case for both $f_v=0.1$ and $f_v=0.2$. This is because the surface adhesiveness makes the particles inclined to aggregate and enhances the forward scattering of radiation. This can be further understood through Fig.\ref{sq} that the structure factor is strongly improved at $q\sim0$ as compared with the HS cases. Therefore the surface stickiness actually plays a competitive role against the random packing (density) effect in forward scattering. In terms of backscattering, both HS and SHS systems show an enhancement over the single particle case. However, for $f_v=0.1$, we find that a higher level of surface stickiness leads to a reduction in backscattering (Fig.\ref{fv01wavl1phasefunc}), while for $f_v=0.2$ the trend is broken (Fig.\ref{fv02wavl1phasefunc}) and the three SHS systems surprisingly exhibit a higher backscattering strength than the HS system. This observation can not be explained by the profile of the structure factor (see Fig.\ref{sq}, $y_{\mathrm{back}}\approx2ka=2.89$ for $\lambda=1\mathrm{\mu m}$).

\begin{figure}[htbp]
	\centering
	\subfloat{	
	\label{fv01C12}
	\includegraphics[width=0.46\linewidth]{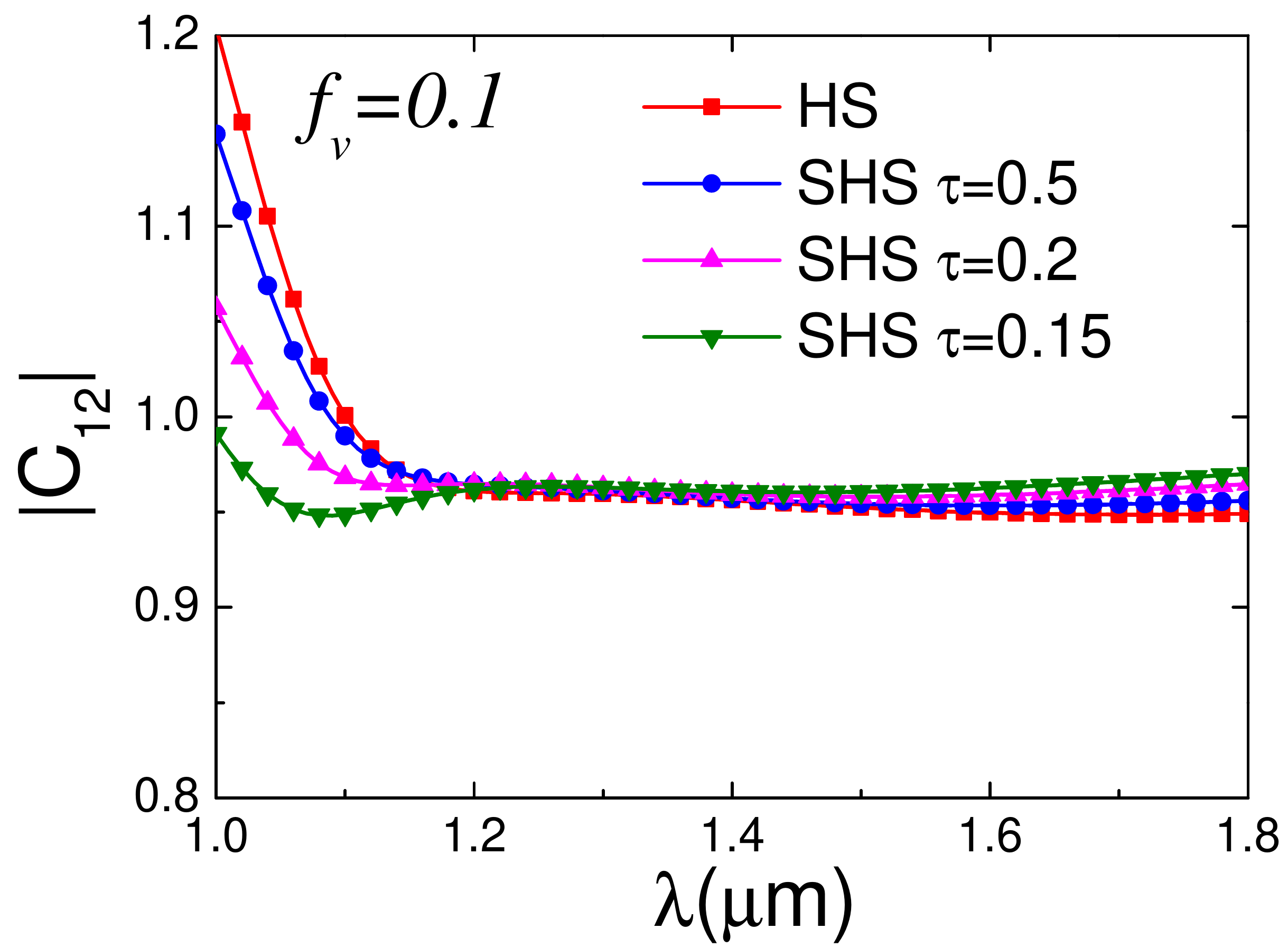}
	
}
    \hspace{0.01in}
    \subfloat{	
	\label{fv01C11}
	\includegraphics[width=0.46\linewidth]{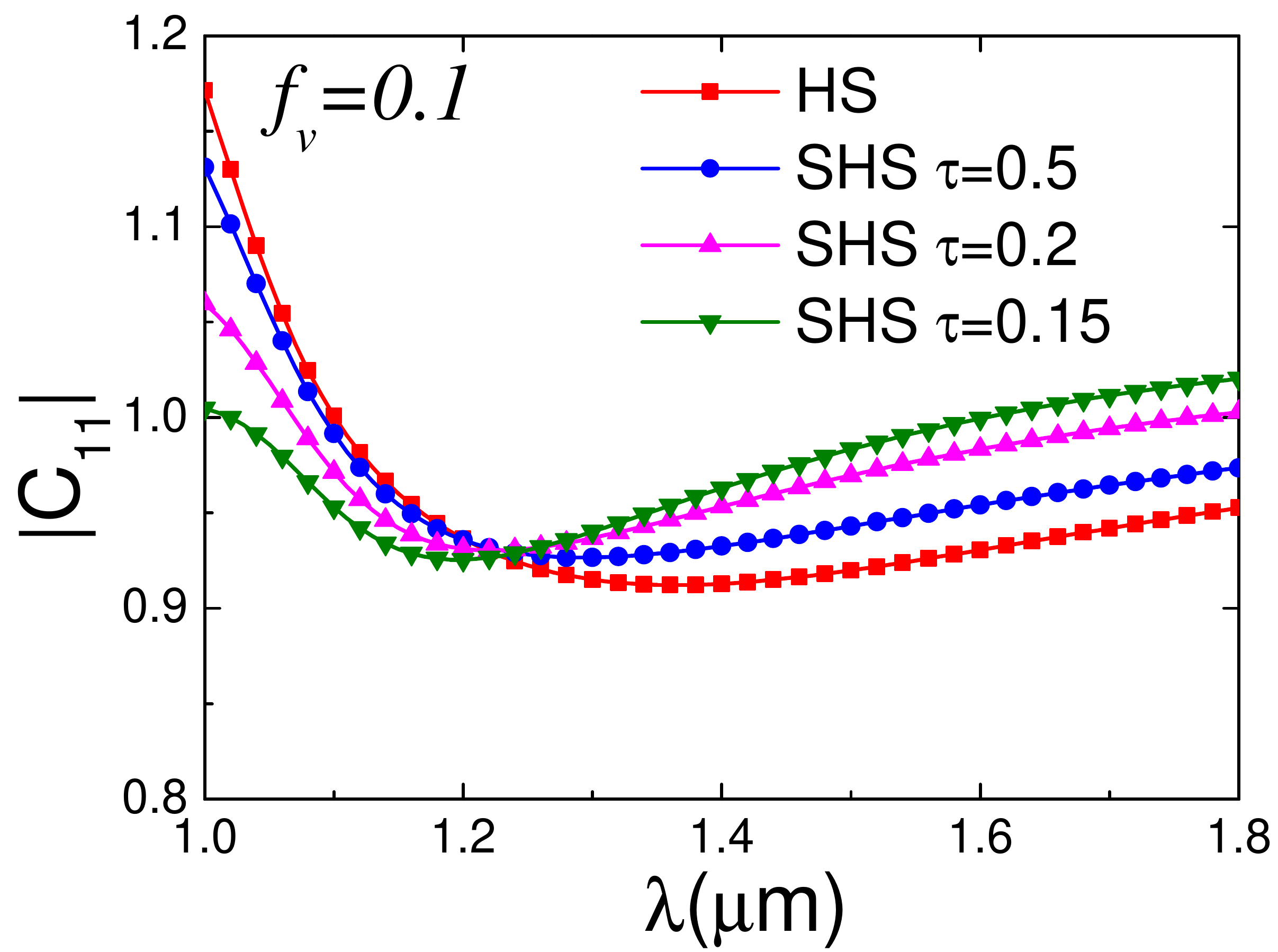}
}
    \hspace{0.01in}
	\subfloat{	
		\label{fv02C12}
		\includegraphics[width=0.46\linewidth]{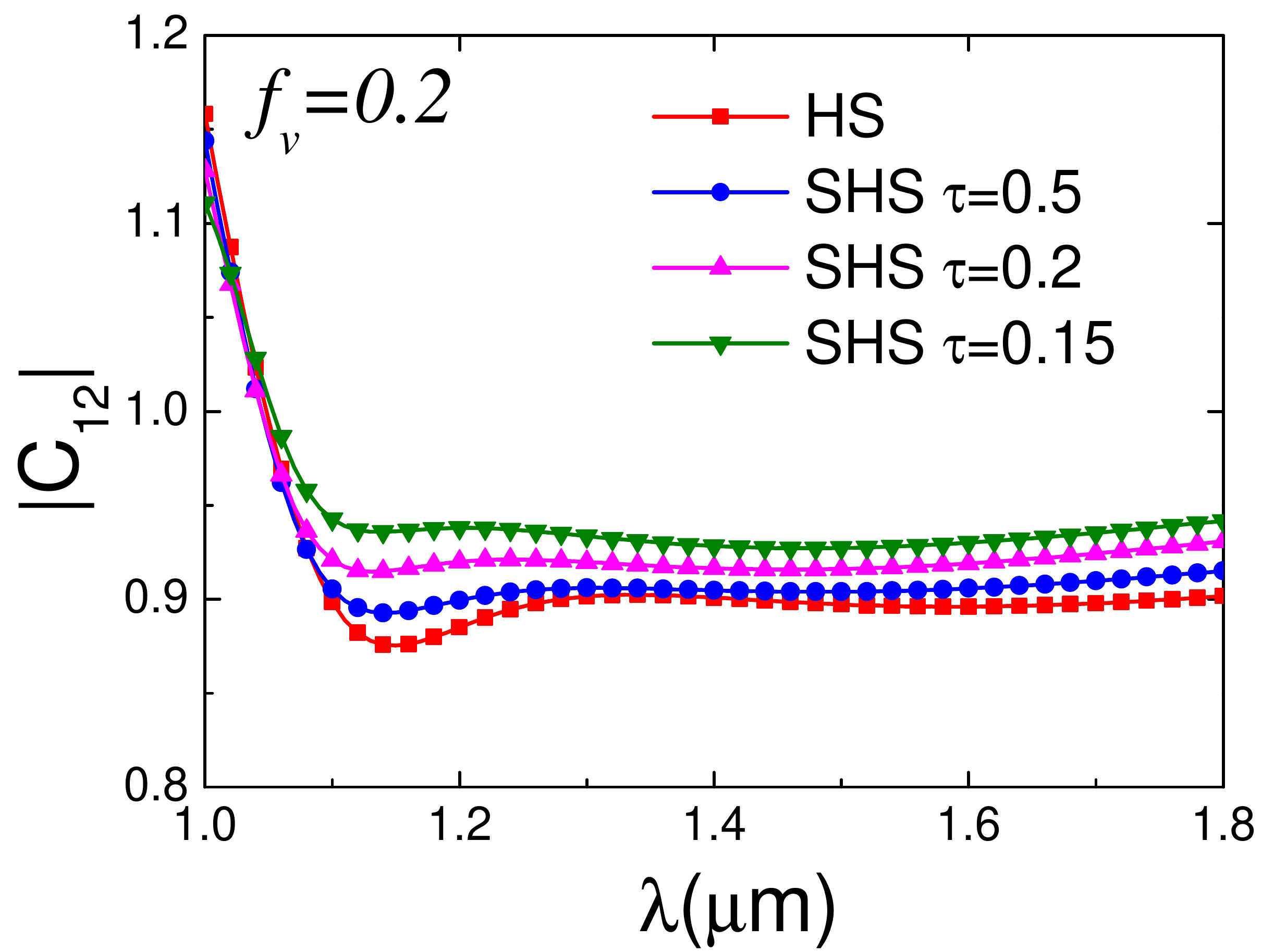}
		
	}
	\hspace{0.01in}
	\subfloat{	
		\label{fv02C11}
		\includegraphics[width=0.46\linewidth]{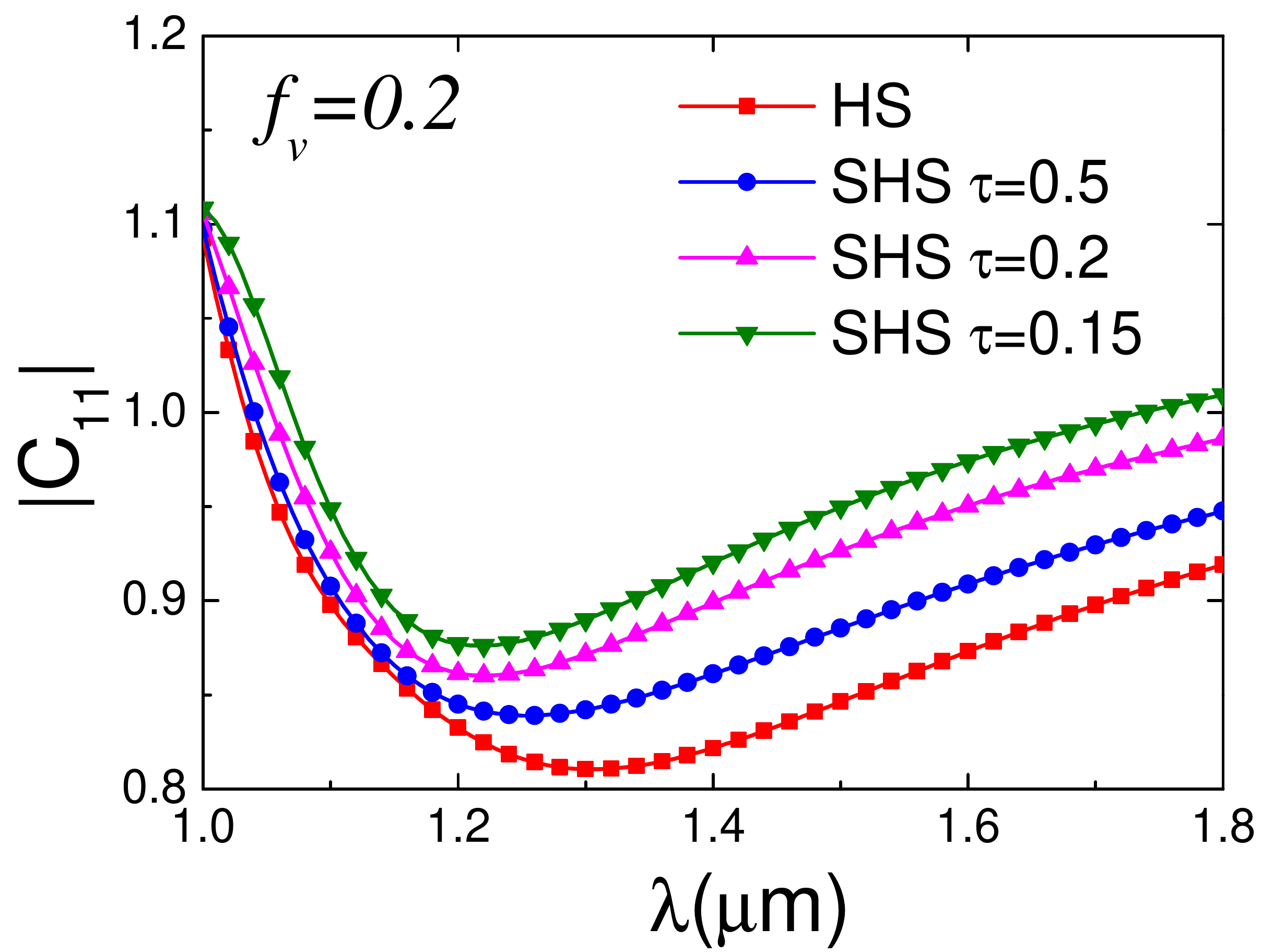}
	}
	\caption{The effective exciting field amplitudes $C_{12}$ (for electric dipole) $C_{11}$ (for magnetic dipole) as a function of the wavelength for random systems with different inverse stickiness parameters. (a) $C_{12}$ for $f_v=0.1$; (b) $C_{11}$ for $f_v=0.1$; (c) $C_{12}$ for $f_v=0.2$; (d) $C_{11}$ for $f_v=0.1$. }\label{fvC}
\end{figure}

As mentioned in Section \ref{theory1}, in the present theory, the structural correlations, manifested as the inverse stickiness parameter $\tau$ (For the HS system, $\tau\rightarrow\infty$ or $\tau^{-1}\rightarrow0$.), affect the asymmetry factor in two ways. The first way is to giving rise to the far-field interference effect as manifested by the structure factor, as analyzed in the above. The second way is to modify the effective electric and magnetic dipole excitations $C_{12}$ and $C_{11}$. Therefore, aiming to find the underlying mechanism of the above observation, we show in Fig.\ref{fvC} the calculated effective exciting field amplitudes for electric and magnetic dipoles for $f_v=0.1$ and $f_v=0.2$ respectively.

It can be observed from Fig.\ref{fvC} that the structural correlations substantially modify the effective exciting field amplitudes, which can be tuned by controlling the inverse stickiness parameter. The structural-correlation-induced effective exciting field amplitudes show a complicated interplay with the single scattering property of the particles. When the single particle is near its scattering resonance ($\lambda\sim 1\mathrm{\mu m}$), $|C|$ is significantly enhanced over unity except for the $\tau=0.15$ case, and increasing the particle stickiness suppresses this enhancement. On the other hand, when a single particle is off-resonance, it is found that increasing the particle stickiness in turn intensifies $|C|$. Note in the case $C_{11}$ for $f_v=0.2$ shown in Fig.\ref{fv02C11}, this transition still exists and is blue-shifted, thus not shown here. This transition can be qualitatively understood as follows. For strongly scattering particles, multiple scattering of electromagnetic waves initially gives rise to a larger exciting field for each particle. However, the surface stickiness makes particles to cluster, leading to a reduction in the exciting field for an individual particle due to the ``screening effect" of other particles against electromagnetic waves \cite{wangPRA2018}. On the contrary, when the particles are weakly scattering, stickiness-induced clustering can further enhance the exciting field for an individual particle, where no screening effect occurs. Another feature can be found from Fig.\ref{fvC} is that the higher packing density (volume fraction) can generally suppress the effective exciting field amplitudes for all inverse sticky parameters. This can be regarded as a screening effect induced by the packing density \cite{wangPRA2018,busch1996PRB}.

To quantitatively demonstrate the role of the effective exciting field amplitudes $C_{12}$ and $C_{11}$ in the asymmetry factor, we plot the effective-field related asymmetry factor $g_\mathrm{C}$ for $f_v=0.1$ and $f_v=0.2$ in Fig.\ref{gcc}, which is defined as follows \cite{Gomez-MedinaPRA2012,wangPRA2018}:
\begin{figure}[htbp]
	\centering
	\subfloat{	
		\label{fv01gcc}
		\includegraphics[width=0.46\linewidth]{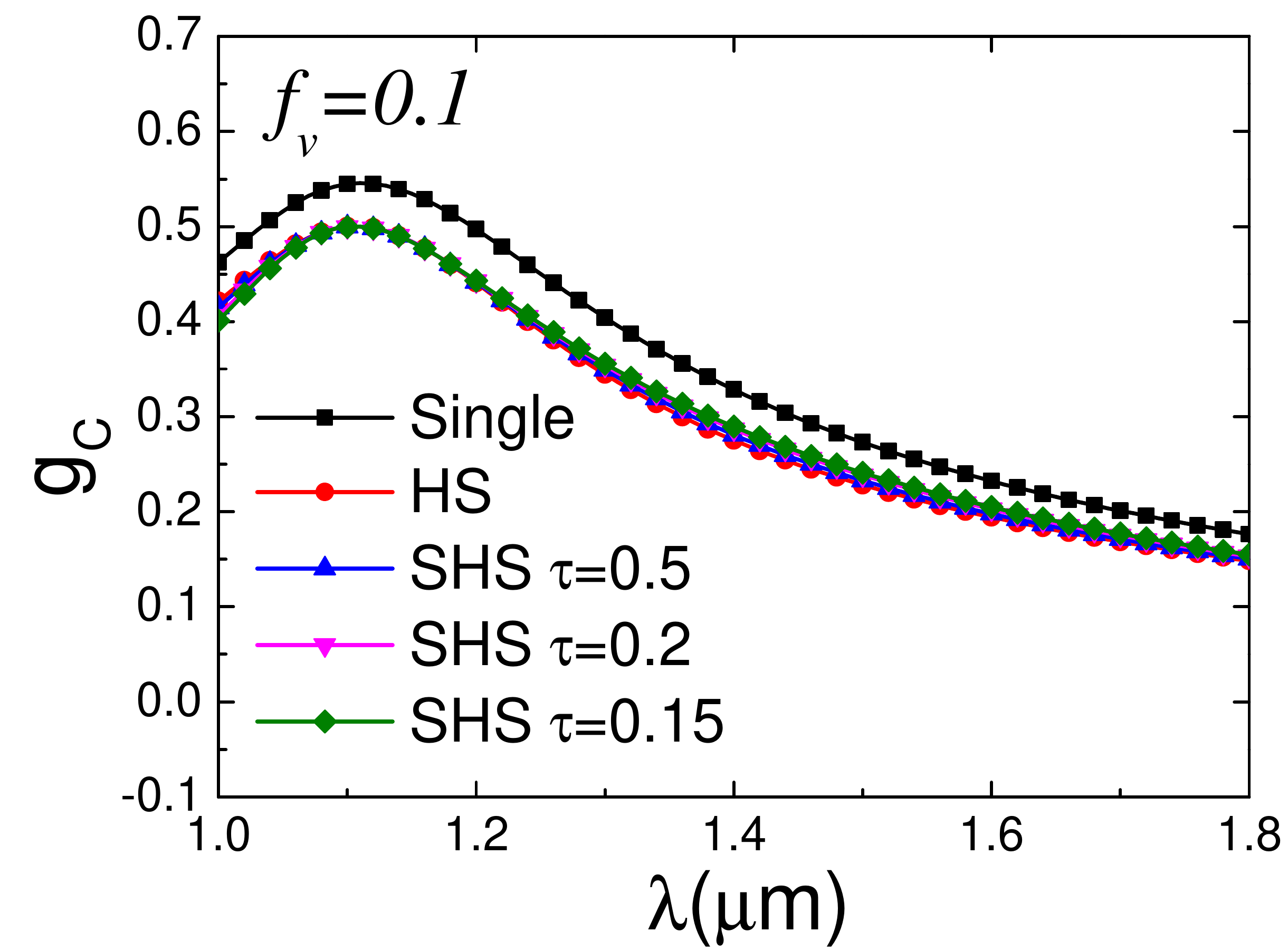}
		
	}
	\hspace{0.01in}
	\subfloat{	
		\label{fv02gcc}
		\includegraphics[width=0.46\linewidth]{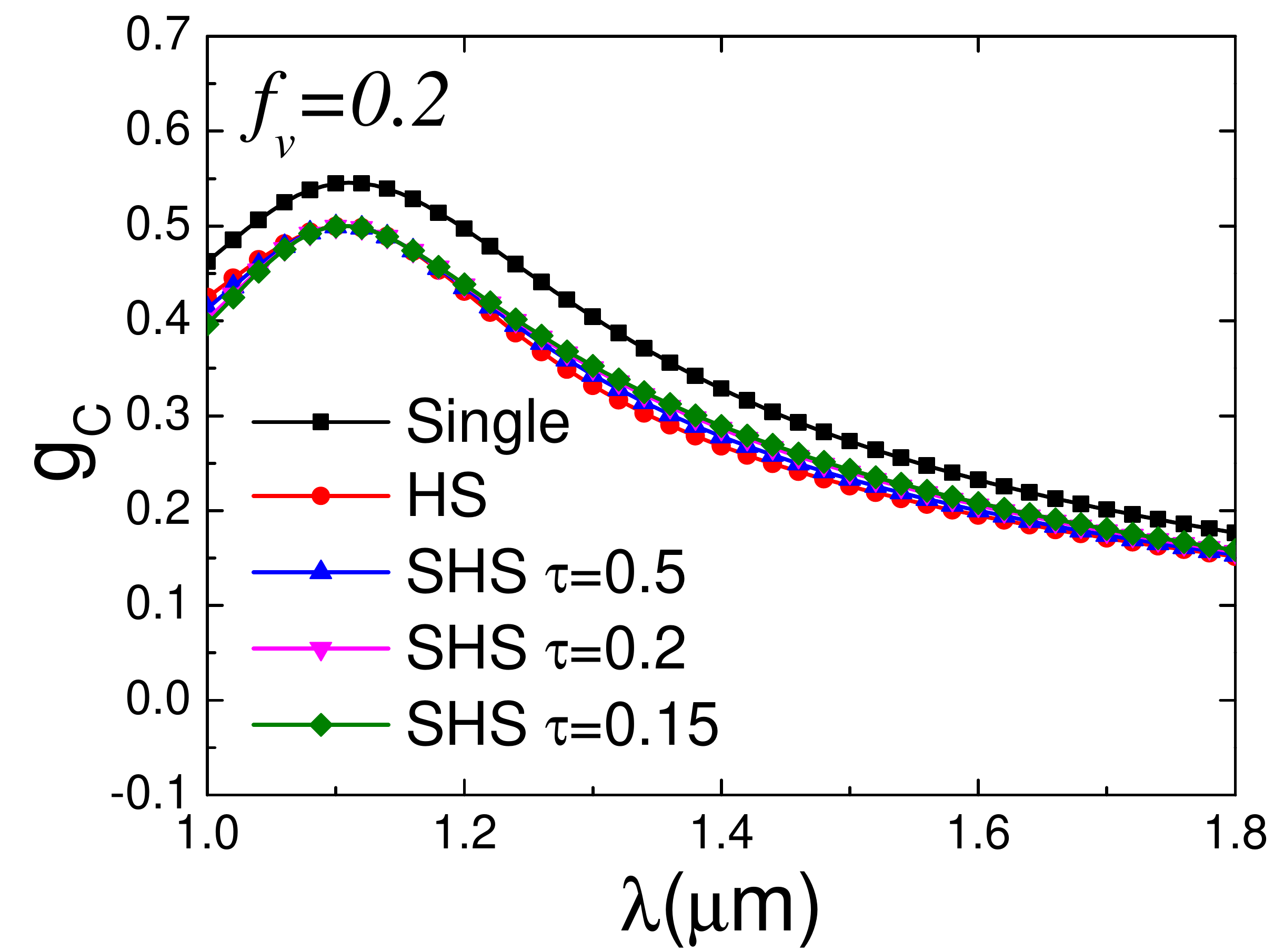}
	}
	\caption{The exciting field related asymmetry factor $g_\mathrm{C}$ for random systems with different stickiness. (a) $f_v=0.1$(b) $f_v=0.2$}\label{gcc}
\end{figure}
\begin{equation}\label{gc}
g_{\text{C}}=\frac{\mathrm{Re}(a_1C_{12}b_1^*C_{11}^*)}{|a_1C_{12}|^2+|b_1C_{11}|^2}.
\end{equation}

\begin{figure}[htbp]
	\centering
	\subfloat{	
		\label{fv01wavl1phasefunc_cc}
		\includegraphics[width=0.46\linewidth]{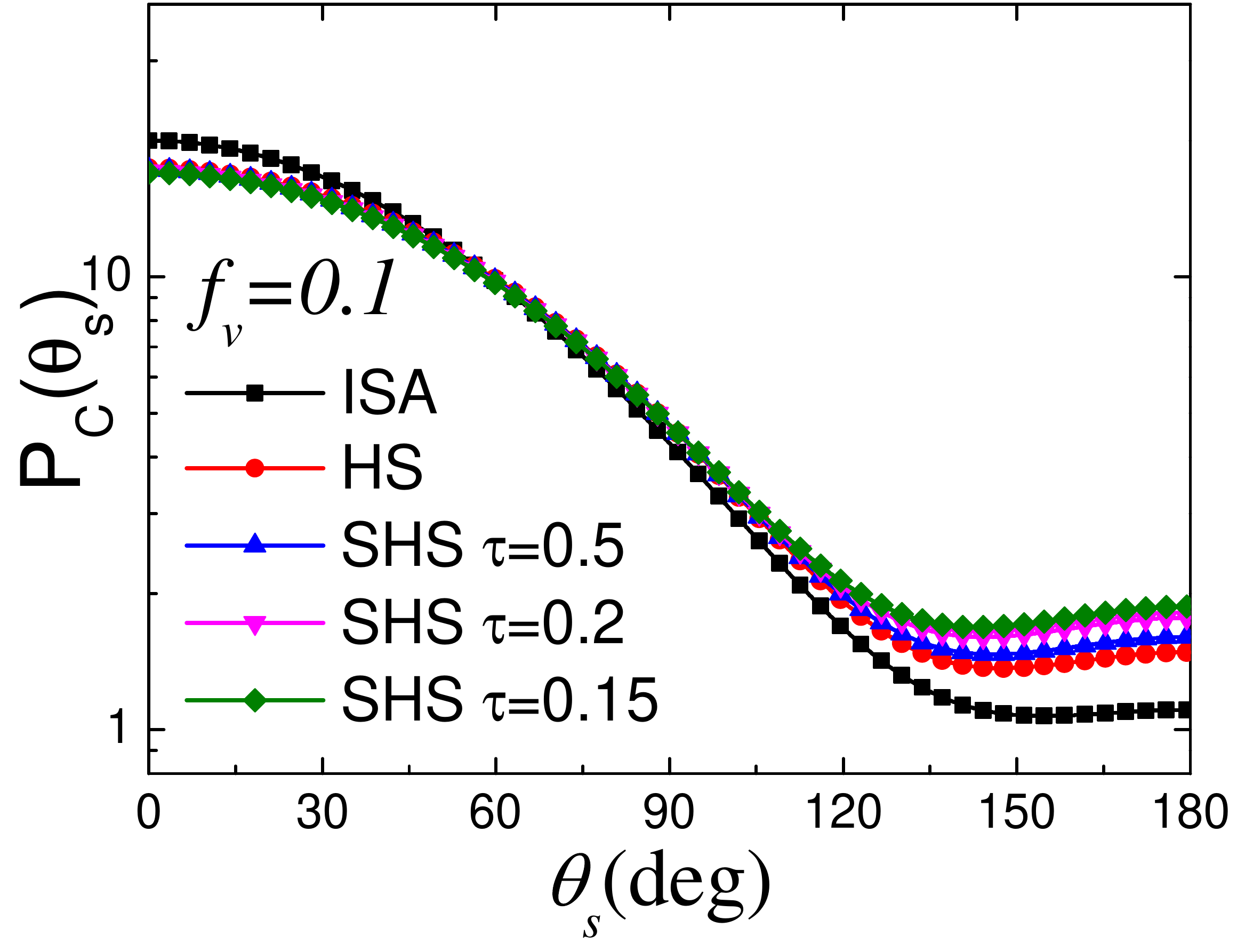}
		
	}
	\hspace{0.01in}
	\subfloat{	
		\label{fv02wavl1phasefunc_cc}
		\includegraphics[width=0.46\linewidth]{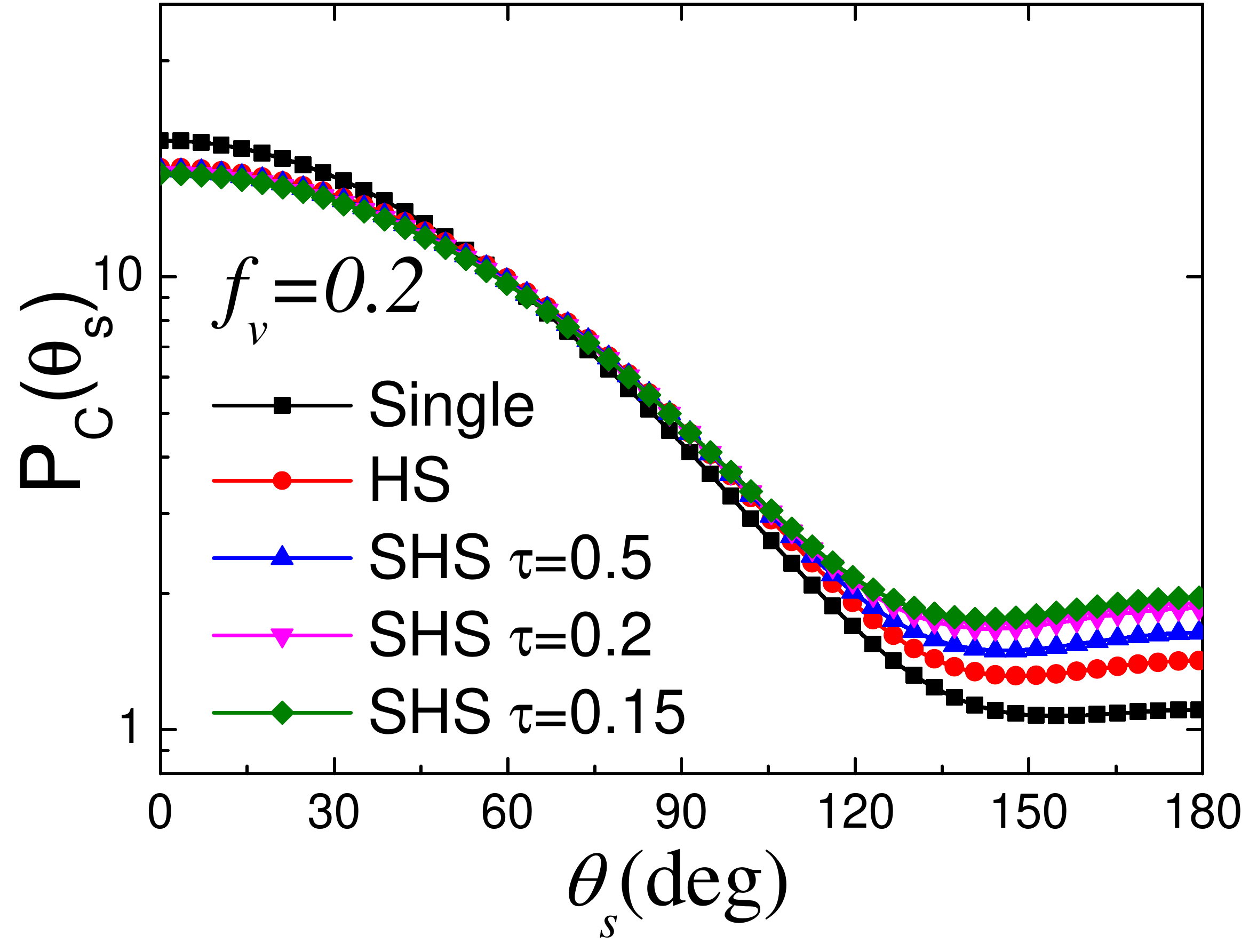}
	}
	\caption{The effective-field related phase function $P_\mathrm{C}(\theta_\mathrm{s})$ for random systems with different stickiness. (a) $f_v=0.1$; (b) $f_v=0.2$.}\label{phasefunc_cc}
\end{figure}
The two subfigures show that when only considering the effect of modification of electric and magnetic dipole excitations, the asymmetry factors for different random systems are also different from that of a single particle. This indicates that this effect indeed plays a role in determining the asymmetry factor of the random media. However, although the effective exciting field amplitudes for various inverse stickiness parameters are substantially different as shown in Fig.\ref{fvC}, their impacts on the obtained $g_\mathrm{C}$ are rather slight. Nevertheless, when considering the detailed phase function, we can find the difference among these random systems. This point can be demonstrated in Fig.\ref{phasefunc_cc} under $\lambda=1\mathrm{\mu m}$. It is shown that the modification of electric and magnetic dipole excitations by different inverse stickiness parameters substantially affects the backscattering probability. However, when increasing volume fraction to $f_v=0.2$, the phase functions change a little, and only the difference among random systems in the backscattering direction is slightly magnified.

\subsection{Scattering coefficient}
\begin{figure}[htbp]
	\centering
	\subfloat{	
		\label{fv01kappas}
		\includegraphics[width=0.46\linewidth]{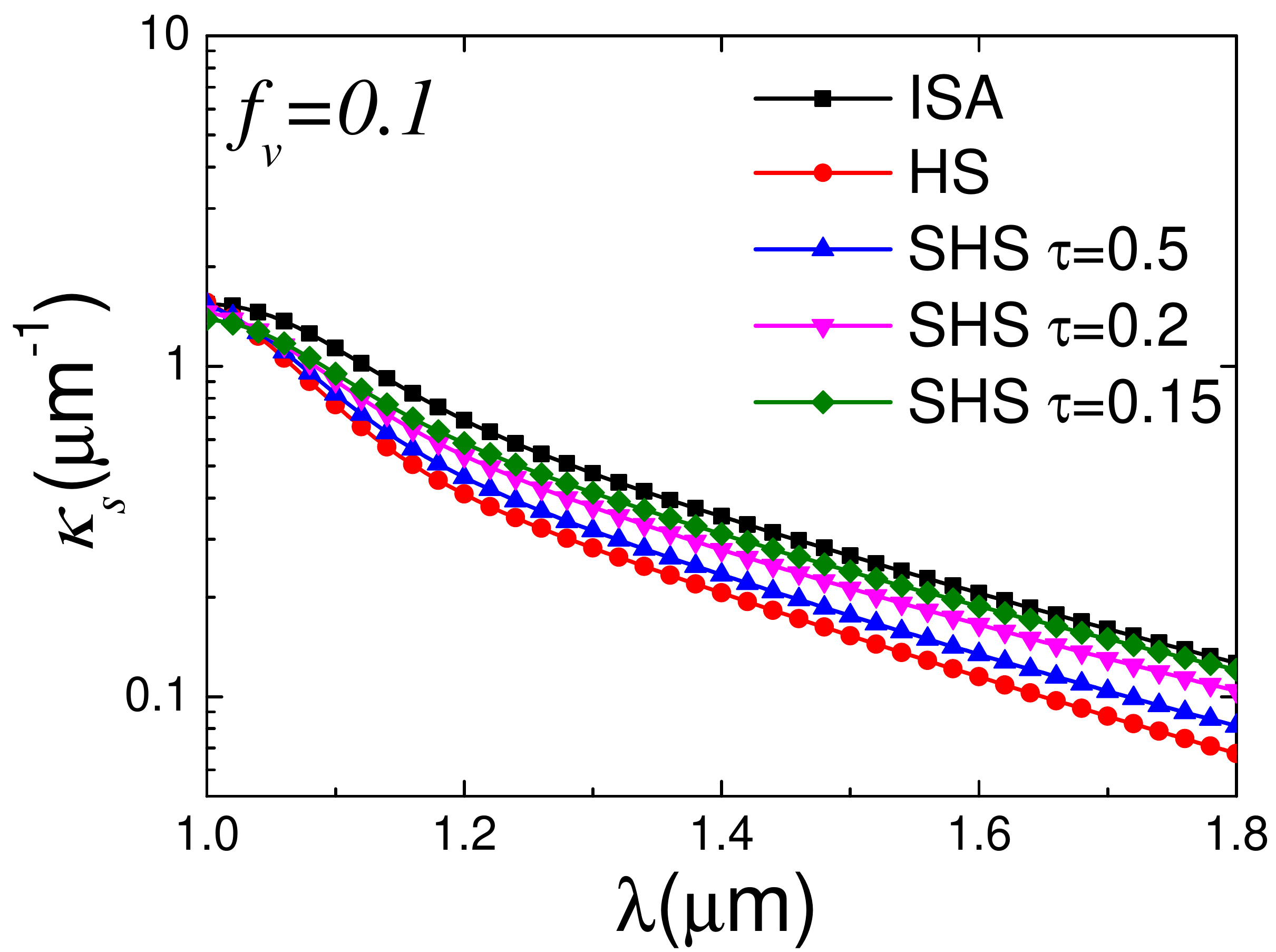}
		
	}
	\hspace{0.01in}
	\subfloat{	
		\label{fv02kappas}
		\includegraphics[width=0.46\linewidth]{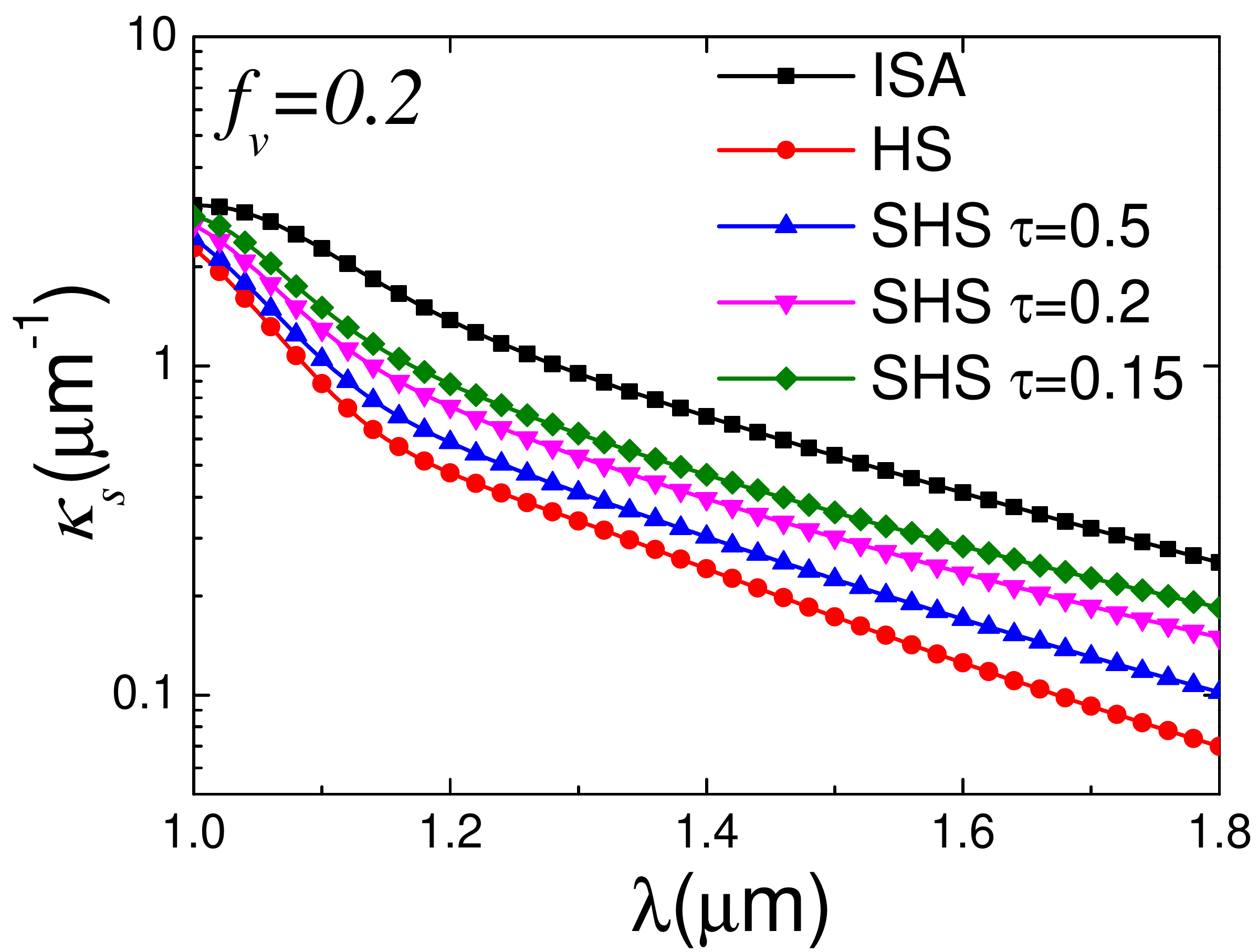}
	}
	\caption{The scattering coefficient $\kappa_s$ for random systems with different stickiness in logarithmic scale. (a) $f_v=0.1$(b) $f_v=0.2$. }\label{kappas}
\end{figure}

In this subsection, we consider how the dependent scattering mechanism affects the scattering coefficient. The scattering coefficient as a function of the thermal radiation wavelength for the particle volume fraction $f_v=0.1$ and $f_v=0.2$ is shown in Figs.\ref{fv01kappas} and \ref{fv02kappas} respectively. 

It can be observed that the different structural correlations affect the scattering coefficient significantly, and the difference is further magnified by the packing density (volume fraction). When the surface stickiness grows, the scattering coefficient is also enhanced over the whole wavelength range, which is due to the enhancement of partial coherence originating from the clustering of the particles \cite{laxPR1952}. This result also provides a way to improve the scattering coefficient by introducing a surface stickiness for the particles.

We further use the method in the previous subsection to analyze the effect of modification of electric and magnetic dipole excitations. We define the following effective-field related scattering coefficient $\kappa_{\text{sC}}$ as:
\begin{equation}\label{kappas_cc_eq}
\begin{split}
\kappa_{\text{sC}}&=\frac{9n_0}{4k^2}\int_{0}^{\pi}
\Big[|a_1C_{21}\pi_n(\cos\theta_\text{s})+b_1C_{11}\tau_n(\cos\theta_\text{s})|^2\\
&+|b_1C_{11}\pi_n(\cos\theta_\text{s})+a_1C_{12}\tau_n(\cos\theta_\text{s})|^2\Big]\sin\theta_\text{s}d\theta_\text{s}.
\end{split}
\end{equation}
\begin{figure}[htbp]
	\centering
	\subfloat{	
		\label{fv01kappas_cc}
		\includegraphics[width=0.46\linewidth]{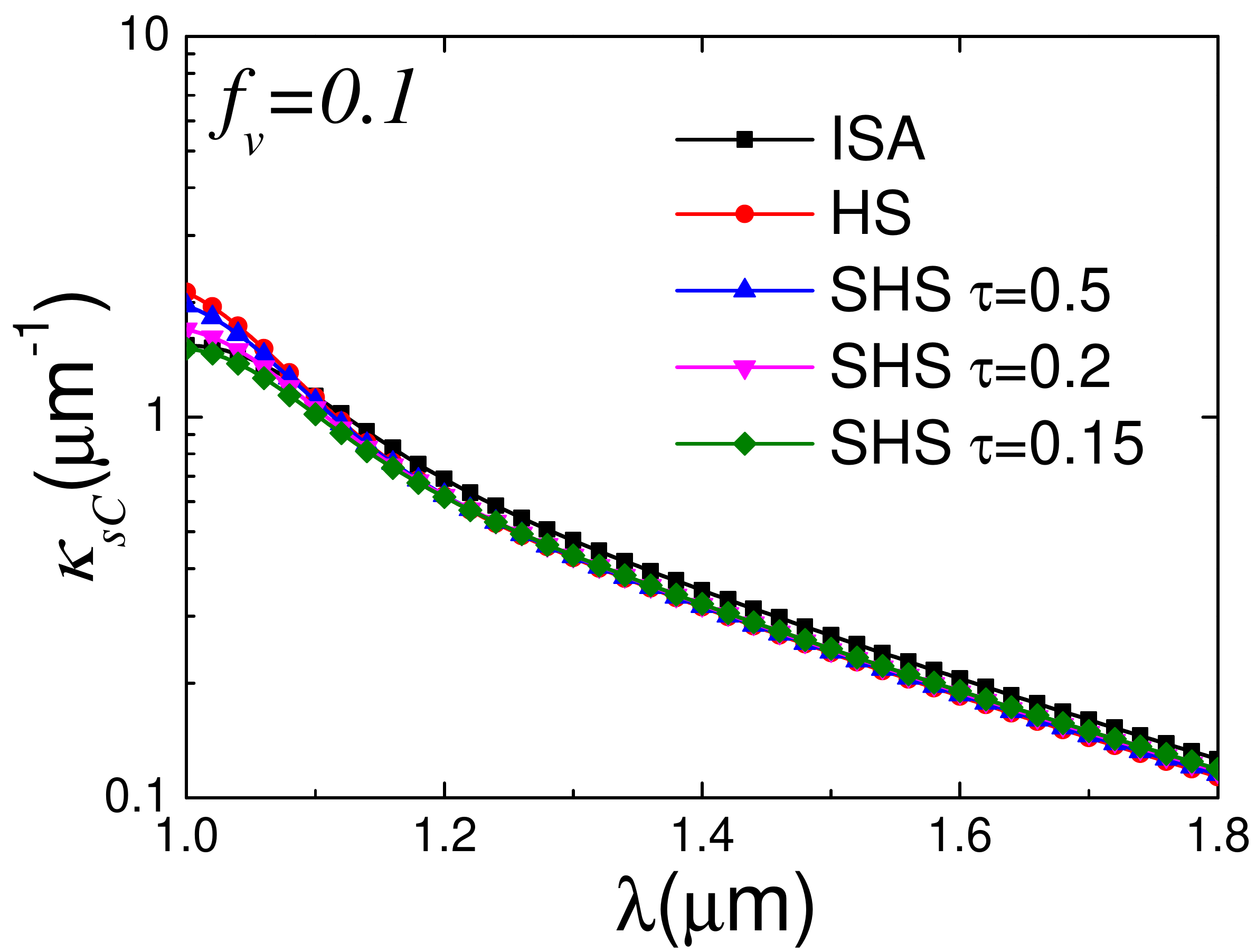}
		
	}
	\hspace{0.01in}
	\subfloat{	
		\label{fv02kappas_cc}
		\includegraphics[width=0.46\linewidth]{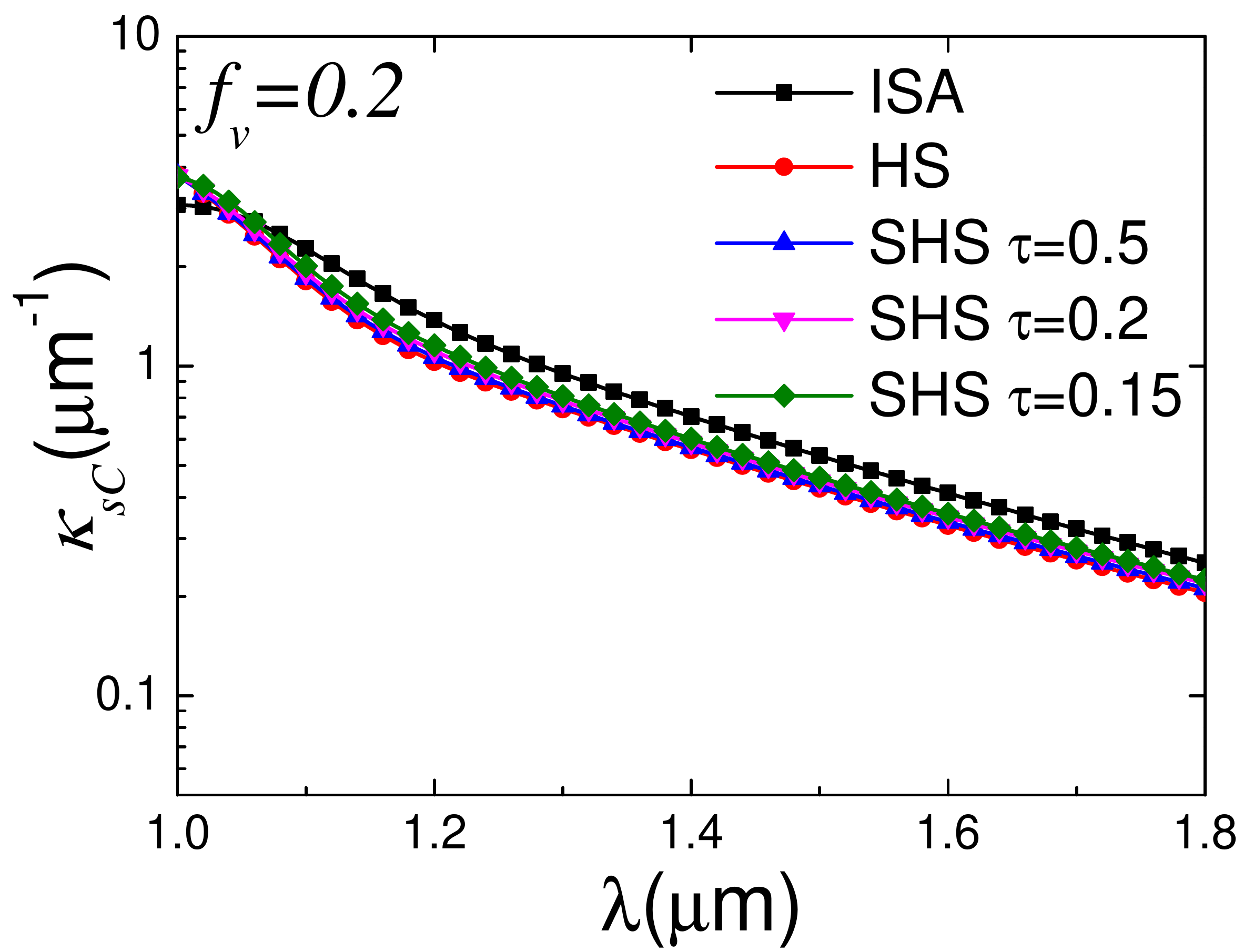}
	}
	\caption{The effective exciting field related scattering coefficient $\kappa_\mathrm{sC}$ for random systems with different stickiness. (a) $f_v=0.1$(b) $f_v=0.2$}\label{kappas_cc}
\end{figure}

The computed results of $\kappa_{\text{sC}}$ for $f_v=0.1$ and $f_v=0.2$ are shown in Fig.\ref{kappas_cc}. Consistent with the results in Figs.\ref{fvC} and \ref{gcc}, the effective exciting field related scattering coefficient indeed exhibits a difference with the result of independent scattering approximation (ISA), and the difference is more significant for the higher packing density, although which is, to some extent, slight. However, the effect of the inverse stickiness parameter is not discriminable. Hence, it can be concluded that it is mainly the structure factor $S(q)$, or the far-field interference effect, that influences the scattering coefficient for the present random media composed of moderate-refractive-index dual-dipolar particles. This result further elucidates the fact that many authors, who only used the ITA model (the far-field interference model) to predict the scattering and transport mean free paths of various random media, could obtained a rather good agreement with their experimental measurements \cite{yamadaJHT1986,kumar1990dependent,fradenPRL1990,mishchenkoJQSRT1994,Peng2007,Liew2011,bresselJSQRT2013,conleyPRL2014,xiaoSciAdv2017}, except for extremely dense media ($f_v>0.4$) \cite{Naraghi2015,Cao1999}. This is, to some extent, a valuable contribution of our present paper.
\section{Conclusions}
In this study, we reveal the role of dependent scattering mechanism on the radiative properties of a random system consisting of dual-dipolar particles. In particular, we investigate the effect of modification of the electric and magnetic dipole excitations and the far-field interference effect, both induced and influenced by the structural correlations. We study in detail how the structural correlations play a role in the dependent scattering mechanism by using two types of particle system, i.e., the hard-sphere system and the sticky-hard-sphere system. We show that the inverse stickiness parameter, which controls the interparticle adhesive force and thus the particle correlations, can tune the radiative properties significantly. Particularly, increasing the surface stickiness can result in a higher scattering coefficient and a larger asymmetry factor. Additionally, the results show that in the present random media composed of moderate-refractive-index dual-dipolar particles, the far-field interference effect plays a dominant role in the radiative properties while the effect of modification of the electric and magnetic dipole excitations is more subtle.
\section*{Acknowledgments}
This work is supported by the National Natural Science Foundation of China (Nos.51636004 and 51476097), Shanghai Key Fundamental Research Grant (16JC1403200) and the Foundation for Innovative Research Groups of the National Natural Science Foundation of China (No.51521004).
\appendix
\section{VSWFs and the translation addition theorem}\label{vswf_appendix}
The regular VSWFs $\mathbf{N}^{(1)}_{mnp}(\mathbf{r})$ for $p=2$ (TE mode) and $p=1$ (TM mode) are defined as \cite{mackowskiJOSAA1996,mackowskiJQSRT2013,tsang2000scattering1,bohrenandhuffman,hulst1957}
\begin{equation}
\mathbf{N}^{(1)}_{mn2}(\mathbf{r})=\sqrt{\frac{(2n+1)(n-m)!}{4\pi n(n+1)(n+m)!}}\nabla\times(\mathbf{r}\psi_{mn}^{(1)}(\mathbf{r})),
\end{equation}
\begin{equation}
\mathbf{N}^{(1)}_{mn1}(\mathbf{r})=\frac{1}{k}\nabla\times\mathbf{N}^{(1)}_{mn2}(\mathbf{r})
\end{equation}
where $k=\omega/c$ is the wave number in free space and $\omega$ is the angular frequency of the electromagnetic wave. $\psi_{mn}^{(1)}(\mathbf{r})$ is regular (type-1) scalar wave function defined as
\begin{equation}
\psi_{mn}^{(1)}(\mathbf{r})=j_n(kr)Y_n^m(\theta,\phi),
\end{equation}
where $j_n(kr)$ is the spherical Bessel function and $Y_n^m(\theta,\phi)$ is spherical harmonics defined as
\begin{equation}
Y_n^m(\theta,\phi)=P_n^m(\cos\theta)\exp(im\phi),
\end{equation}
where we use the convention of quantum mechanics, and $P_n^m(\cos\theta)$ is associated Legendre polynomials.

The outgoing (type-3) VSWFs have can be similarly defined by replacing above spherical Bessel functions with Hankel functions of the first kind $h_n(kr)$.

The translation addition theorem of VSWFs, which transforms the VSWFs centered in $\mathbf{r}_i$ into those centered in $\mathbf{r}_j$, is given by
\begin{equation}
\mathbf{N}^{(3)}_{\mu\nu q}(\mathbf{r}-\mathbf{r}_i)=\sum_{\mu\nu q}A_{mnp\mu\nu q}^{(3)}(\mathbf{r}_i-\mathbf{r}_j)\mathbf{N}^{(1)}_{mnp}(\mathbf{r}-\mathbf{r}_j),
\end{equation}
which is valid for $|\mathbf{r}_i-\mathbf{r}_j|>|\mathbf{r}-\mathbf{r}_j|$, and therefore should be used in the vicinity of $\mathbf{r}_j$. The coefficient $A_{\mu qmp}^{(3)}$ is generally given by \cite{tsang2004scattering2}
\begin{equation}\label{translation_coef3}
\begin{split}
&A_{mn1\mu\nu 1}^{(3)}(\mathbf{r})=A_{mn2\mu\nu 2}^{(3)}(\mathbf{r})=\frac{\gamma_{\mu\nu}}{\gamma_{mn}}(-1)^{m}\\&\cdot\sum_{l}a(\mu,\nu|-m,n|l)a(\nu,n,l)h_l(kr)Y_{l}^{\mu-m}(\theta,\phi),
\end{split}
\end{equation}
\begin{equation}\label{translation_coef4}
\begin{split}
&A_{mn1\mu\nu 2}^{(3)}(\mathbf{r})=A_{mn2\mu \nu 1}^{(3)}(\mathbf{r})=\frac{\gamma_{\mu\nu}}{\gamma_{mn}}(-1)^{m+1}\\&\sum_{l}a(\mu,\nu|-m,n|l,l-1)b(\nu,n,l)h_l(kr)Y_{l}^{\mu-m}(\theta,\phi),
\end{split}
\end{equation}
where $\gamma_{mn}$ is defined as
\begin{equation}
\gamma_{mn}=\sqrt{\frac{(2n+1)(n-m)!}{4\pi n(n+1)(n+m)!}}.
\end{equation}
The coefficients $a(\mu,\nu|-m,n|l)$ and $a(\mu,\nu|-m,n|l,l-1)$ are given by
\begin{equation}
\begin{split}
&a(\mu,\nu|-m,n|l)=(-1)^{\mu-m}\left( 2l+1 \right) \left( \begin{matrix}
\nu&	 n&		l\\
\mu&    -m&		\mu-m\\
\end{matrix} \right) \\&\cdot\left( \begin{matrix}
\nu&	n&		l\\
0&		0&		0\\
\end{matrix} \right)\Big[\frac{(\nu+\mu)!(n-m)!(l-\mu+m)!}{(\nu-\mu)!(n+m)!(l+\mu-m)!}\Big]^{1/2},
\end{split}
\end{equation}
\begin{equation}
\begin{split}
&a(\mu,\nu|-m,n|l,l-1)=(-1)^{\mu-m}\left( 2l+1 \right) \left( \begin{matrix}
\nu&	 n&		l\\
\mu&    -m&		\mu-m\\
\end{matrix} \right) \\&\cdot\left( \begin{matrix}
\nu&	n&		l-1\\
0&		0&		0\\
\end{matrix} \right)\Big[\frac{(\nu+\mu)!(n-m)!(l-\mu+m)!}{(\nu-\mu)!(n+m)!(l+\mu-m)!}\Big]^{1/2},
\end{split}
\end{equation}
in which the variables in the form $\left( \begin{matrix}
j_1&	j_2&		j_3\\
m_1&		m_2&	m_3\\
\end{matrix} \right)$ are Wigner-3$j$ symbols. They can be found in Ref. \cite{abramowitz1964handbook,tsang2004scattering2} and not shown in detail here. Other coefficients $a(\nu,n,l)$ and $b(\nu,n,l)$ are given as \cite{tsang2004scattering2}
\begin{equation}
\begin{split}
&a(\nu,n,l)=\frac{i^{n+l-\nu}}{2n(n+1)}\Big[2n(n+1)(2n+1)+(n+1)(\nu+n\\&-l)
(\nu+l-n+1)-n(\nu+n+l+2)(n+l-\nu+1)\Big],
\end{split}
\end{equation}
\begin{equation}
\begin{split}
&b(\nu,n,l)=-\frac{(2n+1)i^{n+l-\nu}}{2n(n+1)}\Big[(\nu+n+l+1)(n+l-\nu)\\\cdot&
(\nu+l-n)(\nu+n-l+1)\Big]^{1/2}.
\end{split}
\end{equation}
\section{Far-field approximation for outgoing VSWFs}\label{far-field_appendix}

For outgoing (type-3) VSWFs $\mathbf{N}^{(3)}_{mnp}(\mathbf{r}-\mathbf{r}_j)$ centered at $\mathbf{r}_j$, their far-field forms (when $r\gg r_j$) are given by \cite{mackowskiJOSAA1996,tsang2000scattering1,bohrenandhuffman}
\begin{equation}
\begin{split}
\mathbf{N}^{(3)}_{mn2}(\mathbf{r}-\mathbf{r}_j)&\approx i^{-n}\sqrt{\frac{(2n+1)(n-m)!}{4\pi n(n+1)(n+m)!}}\frac{\exp (kr)}{kr}\\&\cdot\exp (-\mathbf{k}_s\cdot\mathbf{r}_j)\mathbf{B}_{mn}(\theta,\phi),
\end{split}
\end{equation}
\begin{equation}
\begin{split}
\mathbf{N}^{(3)}_{mn1}(\mathbf{r}-\mathbf{r}_j)&\approx i^{-n}\sqrt{\frac{(2n+1)(n-m)!}{4\pi n(n+1)(n+m)!}}\frac{\exp(kr)}{kr}\\&\cdot\exp (-\mathbf{k}_s\cdot\mathbf{r}_j)\mathbf{C}_{mn}(\theta,\phi),
\end{split}
\end{equation}
where $\mathbf{B}_{mn}(\theta,\phi)$ and $\mathbf{C}_{mn}(\theta,\phi)$ are vector spherical harmonics. In the present calculation for spheres, only $m=\pm1$ are needed. In this condition,
\begin{equation}
\mathbf{B}_{1n}(\theta,\phi)=-[\hat{\bm{\theta}}\tau_n(\cos\theta)+\hat{\bm{\phi}}\pi_n(\cos\theta)]\exp(i\phi),
\end{equation}
\begin{equation}
\mathbf{B}_{-1n}(\theta,\phi)=\frac{1}{n(n+1)}[\hat{\bm{\theta}}\tau_n(\cos\theta)-\hat{\bm{\phi}}\pi_n(\cos\theta)]\exp(-i\phi),
\end{equation}
\begin{equation}
\mathbf{C}_{1n}(\theta,\phi)=-[\hat{\bm{\theta}}i\pi_n(\cos\theta)-\hat{\bm{\phi}}\tau_n(\cos\theta)]\exp(i\phi),
\end{equation}
\begin{equation}
\mathbf{C}_{-1n}(\theta,\phi)=-\frac{1}{n(n+1)}[\hat{\bm{\theta}}i\pi_n(\cos\theta)+\hat{\bm{\phi}}\tau_n(\cos\theta)]\exp(-i\phi),
\end{equation}
where $\tau_n$ and $\pi_n$ are functions defined as \cite{bohrenandhuffman}
\begin{equation}
\tau_n(\cos\theta)=-\frac{dP_n^1(\cos\theta)}{d\theta},
\end{equation}
\begin{equation}
\pi_n(\cos\theta)=-\frac{P_n^1(\cos\theta)}{\sin\theta}.
\end{equation}
\section*{References}
\bibliography{asym_factor_2018}

\end{document}